\documentclass[journal]{IEEEtran}
\usepackage{xcolor,soul,framed} 
\colorlet{shadecolor}{yellow}
\usepackage[pdftex]{graphicx}
\graphicspath{{../pdf/}{../jpeg/}}
\DeclareGraphicsExtensions{.pdf,.jpeg,.png}
\usepackage[cmex10]{amsmath}
\usepackage{array}
\usepackage{amssymb}
\usepackage{mdwmath}

\usepackage{mdwtab}
\usepackage{eqparbox}
\usepackage{url}
\usepackage{enumerate}
\usepackage{enumitem}
\usepackage{amsfonts}
\usepackage[english]{babel}
\usepackage[utf8]{inputenc}
\usepackage[noend]{algpseudocode}
\usepackage{mathtools}
\usepackage{nomencl}
\usepackage{blindtext}
\usepackage{amsmath}
\usepackage[colorinlistoftodos]{todonotes}
\usepackage{color}
\usepackage{color,soul}
\usepackage{xcolor}
\usepackage{multirow}
\usepackage{graphicx}
\usepackage{epstopdf}
\usepackage[T1]{fontenc}
\usepackage{textcase}
\usepackage{array}
\usepackage{mdwmath}
\usepackage{mdwtab}
\usepackage{eqparbox}
\usepackage{enumerate}
\usepackage{bm}
\usepackage{amsfonts}
\usepackage[ruled,lined,linesnumbered]{algorithm2e}
\usepackage{booktabs}
\usepackage{comment}
\usepackage{amsmath}
\usepackage{mathtools}
\usepackage{caption}
\usepackage{subcaption}
\usepackage{tabularx}
\usepackage{hyperref}
\hypersetup{
    colorlinks=true,
    linkcolor=blue,
    filecolor=magenta,      
    urlcolor=cyan,
}

\usepackage{caption}
\usepackage{mathrsfs}
\usepackage{booktabs}
\usepackage{siunitx}
\usepackage{cite}
\usepackage{url}
\usepackage{setspace}
\usepackage{amsthm}
\usepackage{nomencl}
\makenomenclature
\usepackage{etoolbox}
\renewcommand\nomgroup[1]{%
  \item[\bfseries
  \ifstrequal{#1}{A}{Abbreviations}{%
  \ifstrequal{#1}{V}{Vectors, Matrices, and Sets}{%
  \ifstrequal{#1}{O}{Operators}{%
  \ifstrequal{#1}{S}{Symbols}{}}}}%
]}

\newtheorem{theorem}{Theorem}
\newtheorem{proposition}{Proposition}

\usepackage{xcolor,cite,etoolbox}
\makeatletter 
\pretocmd\@bibitem{\color{black}\csname keycolor#1\endcsname}{}{\fail}

\usepackage{etoolbox}
\makeatletter
\patchcmd{\@makecaption}
  {\scshape}
  {}
  {}
  {}
\makeatother

\title{Blending Data and Physics Against False Data Injection Attack: An Event-Triggered Moving Target Defence Approach}

\author{Wangkun Xu,~\IEEEmembership{Student Member,~IEEE},
       Martin Higgins~\IEEEmembership{Member,~IEEE}, 
       Jianhong Wang,
    Imad M. Jaimoukha, 
    and  Fei Teng,~\IEEEmembership{Senior Member,~IEEE}\\
    \thanks{
    
    Wangkun Xu, Jianhong Wang, Imad M. Jaimoukha, and Fei Teng are with the Department of Electrical and Electronic Engineering, Imperial College London, UK. Martin Higgins is with Department of Engineering Science, University of Oxford, UK. (\textit{Corresponding author: Fei Teng})
    }
}
\begin{document}
\setlength{\textfloatsep}{1pt}
\renewcommand{\baselinestretch}{1}
\markboth{This paper has been accepted by IEEE Trans. on Smart Grid. Copyright of the paper is reserved by IEEE.}%
{Shell \MakeLowercase{\textit{et al.}}: Bare Demo of IEEEtran.cls for IEEE Journals}
\maketitle

\begin{abstract}

Fast and accurate detection of cyberattacks is a key element for a  cyber-resilient  power system. Recently, data-driven detectors and physics-based Moving Target Defences (MTD) have been proposed to detect false data injection (FDI) attacks on state estimation. However, the uncontrollable false positive rate of the data-driven detector and the extra cost of frequent MTD usage limit their wide applications. Few works have explored the overlap between these two areas. To fill this gap, this paper proposes blending data-driven and physics-based approaches to enhance the detection performance. To start, a physics-informed data-driven attack detection and identification algorithm is proposed. Then, an MTD protocol is triggered by the positive alarm from the data-driven detector. The MTD is formulated as a bilevel optimisation to robustly guarantee its effectiveness against the worst-case attack around the identified attack vector. Meanwhile, MTD hiddenness is also improved so that the defence cannot be detected by the attacker. To guarantee feasibility and convergence, the convex two-stage reformulation is derived through duality and linear matrix inequality. The simulation results verify that blending data and physics can achieve extremely high detection rate while simultaneously reducing the false positive rate of the data-driven detector and the extra cost of MTD. All codes are available at \url{https://github.com/xuwkk/DDET-MTD}.


\end{abstract}


\IEEEpeerreviewmaketitle

\mbox{}
\setlength{\nomlabelwidth}{2cm}
\nomenclature[A]{CPSG}{Cyber-Physical Smart Grid}
\nomenclature[A]{BDD}{Bad Data Detection}
\nomenclature[A]{SE}{State Estimation}
\nomenclature[A]{MTD}{Moving Target Defence}
\nomenclature[A]{TPR}{True Positive Rate}
\nomenclature[A]{FPR}{False Positive Rate}
\nomenclature[A]{D-FACTS}{Distributed Flexible AC Transmission System}
\nomenclature[A]{FDI}{False Data Injection}
\nomenclature[A]{OPF}{Optimal Power Flow}
\nomenclature[A]{RTU}{Remote Terminal  Unit}
\nomenclature[A]{DoF}{Degree of Freedom}
\nomenclature[A]{pdf}{Probability Density Function}
\nomenclature[A]{cdf}{Cumulative Density Function}
\nomenclature[A]{LSTM-AE}{Long Short-term Memory AutoEncoder}
\nomenclature[A]{DDET-MTD}{Data-Driven Event-Triggered MTD}
\nomenclature[A]{ROC}{Receiver Operating Characteristic}
\nomenclature[A]{SCADA}{Supervisory Control and Data Acquisition}

\nomenclature[V]{$\bm{O}$, $\bm{I}$}{Zero and identity matrices with appropriate dimensions}
\nomenclature[V]{$\mathbb{R}^n, \mathbb{C}^n$}{Set of real and complex vector}
\nomenclature[V]{$\bm{v}, \widehat{\bm{v}}$}{Actual state and estimated state}
\nomenclature[V]{$\bm{v}', \widehat{\bm{v}}'$}{Actual state and estimated state after MTD}
\nomenclature[V]{$\bm{b}_0, \bm{b}'$}{Susceptance before and after the MTD}
\nomenclature[V]{$\bm{b}^-, \bm{b}^+$}{Lower and upper bounds of susceptance}
\nomenclature[V]{$\bm{z}, \bm{z}_a$}{Normal and attacked measurement}
\nomenclature[V]{$\bm{c}, \bar{\bm{c}}$}{Actual and identified state attack vectors.}
\nomenclature[V]{$\mathcal{B}$}{Susceptance perturbation set}
\nomenclature[V]{$\mathcal{C}$}{Uncertainty set of identified attack vectors}
\nomenclature[V]{$\bm{H}_\star, \bm{S}_\star$}{Jacobian and residual sensitivity matrix of power flow equations w.r.t. $\star$}
\nomenclature[V]{$\bm{H}_\star', \bm{S}_\star'$}{Jacobian and residual sensitivity matrix of power flow equations w.r.t. $\star$ after MTD}
\nomenclature[V]{$\bm{R}$}{Covariance matrix of measurement noise}
\nomenclature[V]{$\bm{R}$}{Covariance matrix of measurement noise}

\nomenclature[O]{$\odot$}{Elemental-wise (Hadamard) product}
\nomenclature[O]{$[\cdot]$}{Diagonalisation on a vector}
\nomenclature[O]{$(\cdot)^*$}{Conjugate on a complex vector}
\nomenclature[O]{$(\cdot)^T$}{Matrix transpose}
\nomenclature[O]{$\text{Col}(\cdot)$, $\text{Ker}(\cdot)$}{Column and kernel space of matrix}
\nomenclature[O]{$\mathcal{P}(\bm{A})$, $\mathcal{P}_{\bm{W}}(\bm{A})$}{Orthogonal and weighted orthogonal projection matrix on $\text{Col}(\bm{A})$ with weight matrix $\bm{W}$}
\nomenclature[O]{$\mathcal{S}(\bm{A})$, $\mathcal{S}_{\bm{W}}(\bm{A})$}{Orthogonal and weighted orthogonal projection matrix on $\text{Ker}(\bm{A}^T)$ with weight matrix $\bm{W}$}
\nomenclature[O]{$\Vert\cdot\Vert_p$}{The $l_p$-norm} 

\nomenclature[S]{$\alpha$}{FPR of BDD}
\nomenclature[S]{$\tau_{\text{lstm}}$}{Threshold of LSTM-AE detector}
\nomenclature[S]{$lr_{\text{detector}}$}{Step size of training LSTM-AE detector}
\nomenclature[S]{$lr_{\text{identifier}}$}{Step size of attack identification}
\nomenclature[S]{$\beta_R, \beta_I$}{Weights of real and imaginary parts of voltage deviations}
\nomenclature[S]{$ite_{\text{min}}, ite_{\text{max}}$}{Maximum and minimum iteration steps in attack identification}
\nomenclature[S]{$\gamma(\bm{z})$}{Residual of measurement $\bm{z}$}
\nomenclature[S]{$M, N+1, P$}{No. of grid branches, buses, and measurements}
\nomenclature[S]{$f(\gamma\vert\kappa)$}{pdf of $\chi^2$ distribution on random variable $\gamma$ with DoF $\kappa$}
\nomenclature[S]{$f(\gamma\vert\kappa, \lambda_c)$}{pdf of non-central $\chi^2$ distribution on random variable $\gamma$ with DoF $\kappa$ and non-centrality parameter $\lambda_c$}
\nomenclature[S]{$\varrho$}{Empirical upper bound on the deviation between the identified and actual attack vectors}
\nomenclature[S]{$\mathbb{P}$}{Symbol of probability}
\nomenclature[S]{$\rho$}{Desired attack detection rate}
\nomenclature[S]{$\lambda(\bm{c}, \bm{b})$}{Non-centrality parameter on attack $\bm{c}$ with susceptance $\bm{b}$}
\nomenclature[S]{$\lambda_c(\rho), \lambda_c$}{Critical non-centrality parameter}
\nomenclature[S]{$\lambda_c'$}{Critical non-centrality parameter with active power flow measurement}

\printnomenclature

\section{Introduction}

\IEEEPARstart{T}{he} Cyber-Physical Smart Grid (CPSG), which is powered by advanced communication and digitalisation techniques, is vulnerable to malicious cyberattacks \cite{mahmoud2021cyberphysical}. Even with limited knowledge, intruders can learn grid information and provide false information based on sensor measurements \cite{khalid2016bayesian}. Recently, False Data Injection (FDI) attack has drawn great attention due to its high stealthiness and adverse impacts on the state estimation (SE) of CPSG \cite{hug2012vulnerability}. The consequences of falsified 
SE include economic losses, transmission line overflow, system instability, and blackout \cite{deng2017false}. To enhance the resilience of the power system, it becomes critical to develop fast and accurately detection for such attacks. The detection algorithms can be broadly classified into model-based and data-driven approaches \cite{bellizio2022transition}. Model-based approaches assume that attackers have imperfect information on the physical model of the power grid and aim to capture the mismatches between real and estimated measurements based on an accurate model\cite{cheng2022highly}. However, the static model information can be targeted and eventually learnt by the attacker \cite{higgins2021topology}, which deteriorates the detection performance. 

\subsection{Model-Based Moving Target Defence}

To overcome the static nature of the model-based detectors, MTD is proposed, in which the system operator can proactively change the reactance of the transmission line through the Distributed Flexible AC Transmission System (D-FACTS) devices \cite{rahman2014moving}. There are three main problems for the design of MTD, namely \textit{`what to move'}, \textit{`how to move'}, and \textit{`when to move'}. Unlike the MTD in information technology (IT) system, the physical structure of the power system and the attack surface should be explicitly considered in CPSG. In detail, `what to move' finds the optimal placement of D-FACTS devices at the planning stage so that the attack surface is minimised \cite{liu2018reactance, zhang2019analysis, liu2020optimal}; `How to move' determines the set-points of the D-FACTS devices during the operation \cite{higgins2022cyber}. For example, \cite{lakshminarayana2020cost} adds the effective constraints on the Optimal Power Flow (OPF) while \cite{liu2022explicit} increases the effectiveness by penalising the cost function. Zhang, \textit{et al.} \cite{zhang2022voltage} considers the voltage stability constrained MTD. Xu, \textit{et al.} \cite{xu2022robust} derives a robust metric to guarantee the effectiveness of MTD on unknown attacks. Recently, \textit{hidden} MTD has been proposed to compete with vigilant attackers who can perform SE and BDD to verify the integrity of grid parameters \cite{tian2018enhanced, zhang2020hiddenness, liu2021optimal, liu2022explicit}. Finally, `when to move' determines the occasion to send the MTD command to the field devices, either \textit{periodically} or \textit{event-triggered}. In most of the literature, MTD is synchronized with SE or OPF, while event-triggered approach is analysed in \cite{higgins2021enhanced, xu2022physical}. 

\subsection{Data-Driven FDI Attack Detector}

Unlike the model-based detector, data-driven approaches do not rely on the model information. Instead, it utilises previous measurements to capture useful spatio-temporal information for detection\cite{wu2021extreme}. To have a good generalization to unseen attack patterns, un-/semi-supervised learning based approaches have been widely researched, such as generative adversarial network \cite{zhang2021detecting} and Long Short-Term Memory AutoEncoder (LSTM-AE) \cite{xu2020deep}. Despite the high detection accuracy, the black-box nature of deep neural networks lacks interpretability, so the detection performance strongly depends on the tuned hyperparameters in the training data set \cite{pang2021deep}. As a result, high False Positive Rate (FPR) on the unseen normal measurements becomes one of the fundamental challenges in applying data-driven detectors \cite{ahmed2020challenges}. This trade-off is reported in \cite{ashok2018online} where the forecast-aided detector suffers from 20\% FPR to achieve 90\% True Positive Rate (TPR) for attacks with small strength. 



\subsection{Contributions}

Both MTD and data-driven FDI attack detectors overlook the rarity of the attack and can deteriorate the normal operation to some extent. In detail, the cumulative additional cost of frequent use of MTD is significant. Meanwhile, as data availability is of high priority in CPSG, continual false alarms from a data-driven detector cause frequent contingencies and overload response resources, compromising the operator's confidence in the detector. Table \ref{tab:compare_mtd_data} compares the MTD and data-driven detector, showing a clear complementation on each other. Therefore, we consider using interpretable physics-based MTD to verify the decision from data-driven detector, which in turn serves as an event triggering on the MTD to reduce the operational cost. The proposed \textit{Data-Driven Event-Triggered MTD} (DDET-MTD) framework can achieve high TPR, low FPR, less operation cost, and great interpretability. 

\begin{table}[h]
    \centering
    \footnotesize
    \caption{Comparison on MTD and data-driven detector}
    \begin{tabularx}{\linewidth}{l|X|X}
    \hline\hline
         &  \textbf{Advantages} & \textbf{Disadvantages} \\\hline
        \textbf{MTD} & High interpretability; Controllable FPR  & High operation cost for frequent implementation \\\hline
        \textbf{Data-Driven} & Fast response; No extra operation cost & Low interpretability; Uncontrollable FPR \\\hline\hline
    \end{tabularx}
    \label{tab:compare_mtd_data}
\end{table}

The contributions are highlighted as follows.


\begin{enumerate}
    \item For the first time, a novel event-triggering framework is proposed that seamlessly links the design and implementation of a data-driven detector and physics-based MTD. The proposed framework outperforms the individual approach by rejecting false positive decisions from the data-driven detector and reducing the use and cost of MTD. 
    \item A novel measurement recovery algorithm is proposed to identify attacks through normality projection. The FDI attack detector and identifier are integrated into a single LSTM-AE deep learning model, while power system physics information is embedded to ensure the fidelity of the recovered attack. 
    \item A bilevel optimisation problem is formulated for the MTD design. In the upper level, hiddenness is improved while in the lower level, the detection accuracy is robustly guaranteed on the worst-case attack around the identified attack vector. To guarantee the feasibility and the convergence, the nonlinear nonconvex bilevel optimisation is further relaxed into two successive semidefinite programmings using linear matrix inequalities and duality.
    \item The performance of the algorithm is verified with two benchmark algorithms under periodic and event trigger settings, using real-time load and solar profiles.
    
\end{enumerate}




The remainder of the paper is organised as follows. Preliminaries are given in Section \ref{sec:preliminary}. The proposed DDET-MTD algorithms are described in Section \ref{sec:data_mtd}. The results are analysed in Section \ref{sec:simulation} and this paper concludes in Section \ref{sec:conclusion}. Additional material and proofs can be found in the appendix.

\section{Preliminaries}\label{sec:preliminary}

In this section, the power system model, state estimation, and FDI attacks are reviewed. Two detection algorithms, the LSTM-AE detector and MTD, are also introduced.


\subsection{System Model}

The power system can be modelled as a graph $\mathcal{G}(\mathcal{N},\mathcal{E})$ where $\mathcal{N}$ and $\mathcal{E}$ are the sets of buses and branches with numbers $|\mathcal{N}| = N+1$ and $|\mathcal{E}| = M$, respectively. The complex voltage on the bus $n$ is indicated as ${\bm{v}_n}=|\bm{v}_n|\angle\bm{\theta}_n$; and the admittance on the branch $m$ is indicated as $\bm{y}_m=\bm{g}_m+\mathrm{j}\bm{b}_m$. Let $\bm{z}\in\mathbb{R}^{P}$ be the vector of measurements; the power measurement equation can be written as 
\begin{equation}\label{eq:measure_nonlinear_1}
    \bm{z}=\bm{h}(\bm{x})+\bm{e}    
\end{equation}
where $\bm{e}$ is the zero-mean Gaussian measurement error with diagonal covariance matrix $\bm{R} = \text{diag}[\sigma_1^2,\sigma_2^2,\dots,\sigma_m^2]$. In this paper, the RTU measurements are considered as follows \cite{zimmerman2018matpower}:

1). Complex power injections:
\begin{equation*}
    \bm{S}_{bus} = [\bm{v}]\bm{Y}_{bus}\bm{v}^*
\end{equation*}

2). `from' and `to'-side complex power flows:
\begin{subequations}
    \begin{equation*}
        \bm{S}_f = [\bm{C}_f\bm{v}]\bm{Y}_f^*\bm{v}^*
    \end{equation*}
     \begin{equation*}
        \bm{S}_t = [\bm{C}_t\bm{v}]\bm{Y}_t^*\bm{v}^*
     \end{equation*}
\end{subequations}
where $\bm{v}\in\mathbb{C}^{N+1}$ is vector of complex bus voltages; $\bm{C}_f$ and $\bm{C}_t\in\mathbb{R}^{M\times (N+1)}$ are the `from' and `to' side incidence matrices, respectively; $\bm{Y}_{bus}\in \mathbb{R}^{(N+1) \times (N+1)}$ is the bus admittance matrix; $\bm{Y}_f$ and $\bm{Y}_t \in\mathbb{R}^{M\times (N+1)}$ are the `from' and `to' side branch incidence matrices, respectively. The total measurement becomes $\bm{z}=[\bm{P}_{bus}^T, \bm{P}_f^T, \bm{P}_t^T, \bm{Q}_{bus}^T, \bm{Q}_f^T, \bm{Q}_t^T]^T\in\mathbb{R}^{2N+4M+2}$. Detailed formulations of $\bm{Y}_{bus}$, $\bm{Y}_f$, and $\bm{Y}_t$ can be found in \cite{zimmerman2018matpower}.

Given redundant measurements, the power system SE acquires voltage phasors at all buses by solving the weighted least-square problem using the Gauss-Newton algorithm \cite{abur2004power}:
\begin{equation}\label{eq:se_nonlinear}
    \widehat{\bm{v}} =  \arg\min_{{\bm{v}}} J({\bm{v}})=(\bm{z}-\bm{h}({\bm{v}}))^{T} \bm{R}^{-1} (\bm{z}-\bm{h}({\bm{v}}))
\end{equation}
where $\widehat{\bm{v}}$ is the estimated state. 

Based on the estimated state, the Bad Data Detection (BDD) raises an alarm if the measurement residual is higher than a predefined threshold \cite{abur2004power}. Given $\bm{\hat{\nu}}$, the residual vector can be written as the difference between the observed and estimated measurements $\bm{r} = \bm{z}-\bm{h}(\bm{\hat{\nu}})$ and the residual can be represented as $\gamma(\bm{z}) = \|\bm{R}^{-\frac{1}{2}}\bm{r}\|_2^2$. Since $\bm{R}^{-\frac{1}{2}}\bm{r}$ approximately follows standard normal distribution and at least $2N$ measurements have to be observed to solve \eqref{eq:se_nonlinear}, $\gamma(\bm{z})$ approximately follows the $\chi^2$ distribution with DoF $\kappa = P-2N$ \cite{abur2004power}.
Therefore, letting $f_\chi(\gamma|\kappa)$ represent the density function of $\gamma$, the system operator can decide the detection threshold $\tau_\chi(\alpha)$ by a tolerable FPR $\alpha$ such that 
\begin{equation}\label{eq:chi_square_1}
    \mathbb{P}(\gamma \geq \alpha) = \int_{\tau_\chi(\alpha)}^\infty f_\chi(\gamma|\kappa)du = \alpha
\end{equation}
where the typical value of $\alpha$ is $1\%-5\%$. 

\subsection{FDI Attacks}

Given the measurement $\bm{z}$, an attacker can launch FDI attacks by formulating $\bm{z}_a = \bm{z}+\bm{a},~\bm{a}\in\mathbb{R}^{P}$, which cannot be detected by the BDD if $\bm{a} = \bm{h}(\widehat{\bm{v}}+\bm{c}) - \bm{h}(\widehat{\bm{v}})$ \cite{hug2012vulnerability}. Under this condition, $\bm{z}_a = \bm{h}(\bm{v}+\bm{c}) + \bm{e}$ whose residual is unchanged as \eqref{eq:chi_square_1}. To successfully launch FDI attacks, the attacker's abilities are assumed as follows:

\textbf{Assumption One}. The attackers are aware of the topology and parameters of the grid to build $\bm{h}(\cdot)$, which can be circumvented by data-driven algorithms \cite{higgins2021topology}. However, the data collection is time-consuming, e.g. several hours \cite{lakshminarayana2020cost}.

\textbf{Assumption Two}. The attacker can access and modify all sensor measurements. This can be achieved by hijacking all RTU measurements or by changing the Domain Name Systems server between the SCADA front end and control centre \cite{zhang2021smart}. Meanwhile, the attack strength is limited because the attacked state should be within the normal range \cite{gao2022novel}.

\textbf{Assumption Three}. The attacker can verify his knowledge about the grid parameters by checking the integrity of the hijacked measurement. Similarly to BDD, the attacker can perform SE, and if the residual is greater than the threshold, the attacker will not carry out the attack but will turn to collecting more information \cite{tian2018enhanced}.

Assumption One to Three require the attacker's effort to gain accurate gird topology and parameters, which may not be practical in real-time operation. However,  we assume the strongest attacker and study the general defence algorithm against the unpredictable attacker, which is in line with the assumptions made in \cite{liu2018reactance, zhang2019analysis, liu2020optimal, lakshminarayana2020cost, liu2022explicit}.

\subsection{LSTM-AE based Data-Driven Detector}

\label{para:lstm-ae}

\begin{figure}[]
     \centering
     \includegraphics[width=0.98\linewidth]{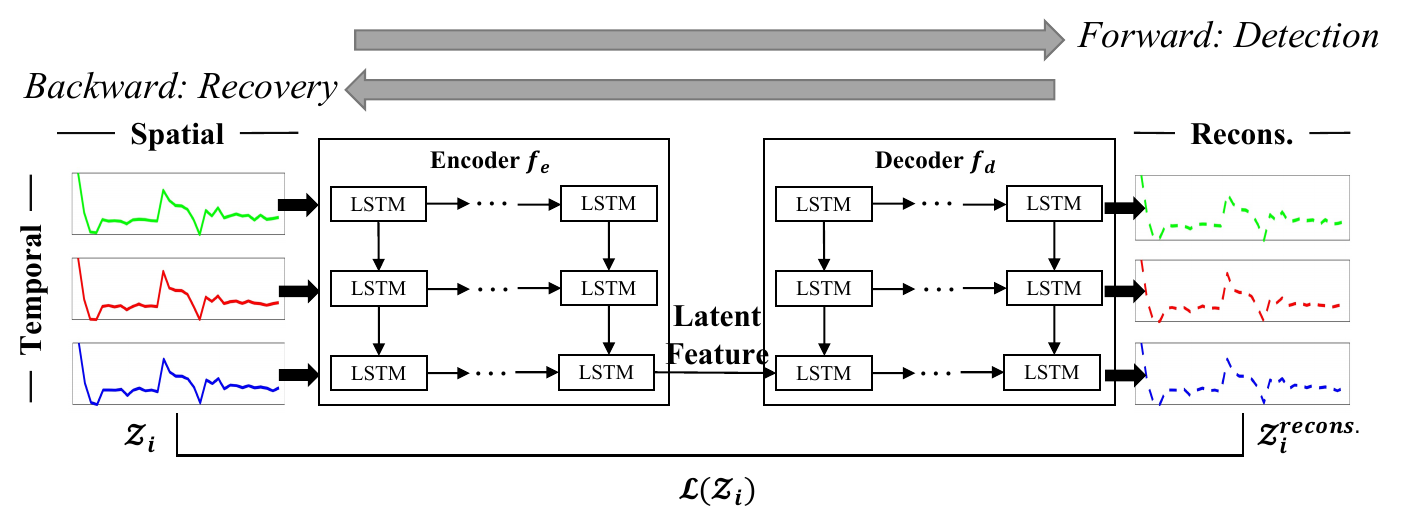}
     \caption{An illustrative structure of LSTM-AE.}
     \label{fig:lstm_ae}
\end{figure}

Although the attacker can launch FDI attacks by exploring the grid topology and parameters, the attacked measurement $\bm{z}_a^{<t>}$ at time $t$ may violate the trend in a certain time window of length $T$. Our previous work in \cite{xu2020deep} designed a \textit{semi-supervised} data-driven detector using LSTM-AE to explicitly learn the \textit{spatiotemporal} correlations in sequential measurements. Fig. \ref{fig:lstm_ae} illustrates the structure of LSTM-AE where each column of connected LSTM cells represents one layer of the deep recurrent network. 
Given a set of $L$ normal measurements $\mathcal{Z}=\{\bm{z}^{<1>},\bm{z}^{<2>},\dots,\bm{z}^{<L>}\}$, consider a length-$T$ continuous subset $\mathcal{Z}_i = \{\bm{z}^{<t_i>},\bm{z}^{<t_i+1>},\cdots, \bm{z}^{<t_i+T-1>} \}$. At each layer, LSTM cells contain `states' whose values depends on the previous memories and can be updated or forgotten by the current measurement. To learn the temporal pattern of the measurements, the LSTM-AE is trained to compress its input $\mathcal{Z}_i$ into a latent representation of the lower dimension, while only normal data can be successfully recovered by the decoder; thus, real-time attack measurements can be distinguished by directly evaluating the loss function:
\begin{equation}\label{eq:lstm_loss}
    \mathcal{L}(\mathcal{Z}_i)=\frac{1}{T P}\sum_{j=0}^{T-1}\left\|{\bm{z}}^{<t_i+j>}-f_{d}\left(f_{e}\left({\bm{z}}^{<t_i+j>} \right)\right)\right\|_{2}^{2}
\end{equation}
where $f_e$ and $f_d$ represent the encoder and decoder mappings respectively. The detection threshold $\tau_{\text{lstm}}$ can be defined based on the distribution of the residual $\mathcal{L}(\mathcal{Z}_i)$ in the validation set \cite{xu2020deep}. Although the attacker may also exploit the temporal correlations between the measurements, we assume that they cannot know the exact temporal pattern learnt from the LSTM-AE detector.

\subsection{Moving Target Defence}

Compared to the data-driven detector, model-based detections are more likely accepted by the system operator due to its high interpretability. To overcome the static nature of the model-based detector, MTD is introduced to proactively change the grid parameters using D-FACTS devices. The typical reactance perturbation ratio is less than 50\% \cite{ lakshminarayana2020cost}. For convenience, the constraint on the reactance is converted to the constraint on the susceptance as follows.
\begin{equation*}
    \bm{h}_{\bm{b}_0}(\cdot)\xrightarrow{\text{MTD}} \bm{h}_{\bm{b}'}(\cdot)
\end{equation*}
where $\bm{b}' = \bm{b}_0 + \Delta \bm{b}$ are the susceptances after activating the D-FACTS devices. Details on the reactance to susceptance conversion can be found in Appendix \ref{sec:app_susceptance}. Physical constraints can be represented by the set $\mathcal{B}=\{\bm{b}'|\bm{b}^-\leq\bm{b}'\leq\bm{b}^+\}$ where $\bm{b}^{-}$ and $\bm{b}^{+}$ are the lower and upper bound of the susceptance. If there is no D-FACTS device in branch $i$, $\bm{b}^-_i=\bm{b}^+_i=\bm{b}_{0i}$.

If there is no attack, the post-MTD measurement still follows the $\chi^2$ distribution. Therefore, no additional FPR is introduced by MTD. In contrast, if the attack exists, the residual vector will no longer follow the $\chi^2$ distribution of the legitimate measurement and hence trigger the BDD alarm. In detail, \textit{MTD effectiveness} refers to the accuracy of BDD after MTD is activated \cite{liu2018reactance}. Recent literature also proposes the concept of \textit{MTD hiddenness} by noticing that the prudent attacker can also check the integrity of model parameters using BDD-like method \cite{tian2018enhanced}. According to Assumption Three, the system will therefore face new threats \cite{tian2018enhanced}. Apart from achieving high detection rate, the hidden MTD requires reducing the attacker's residual so that the attackers keeps using out-of-date grid knowledge to formulate the attack. 

\section{Data Driven Event-Triggered MTD}
\label{sec:data_mtd}

\begin{figure*}[]
     \centering
     \includegraphics[width=0.78\linewidth]{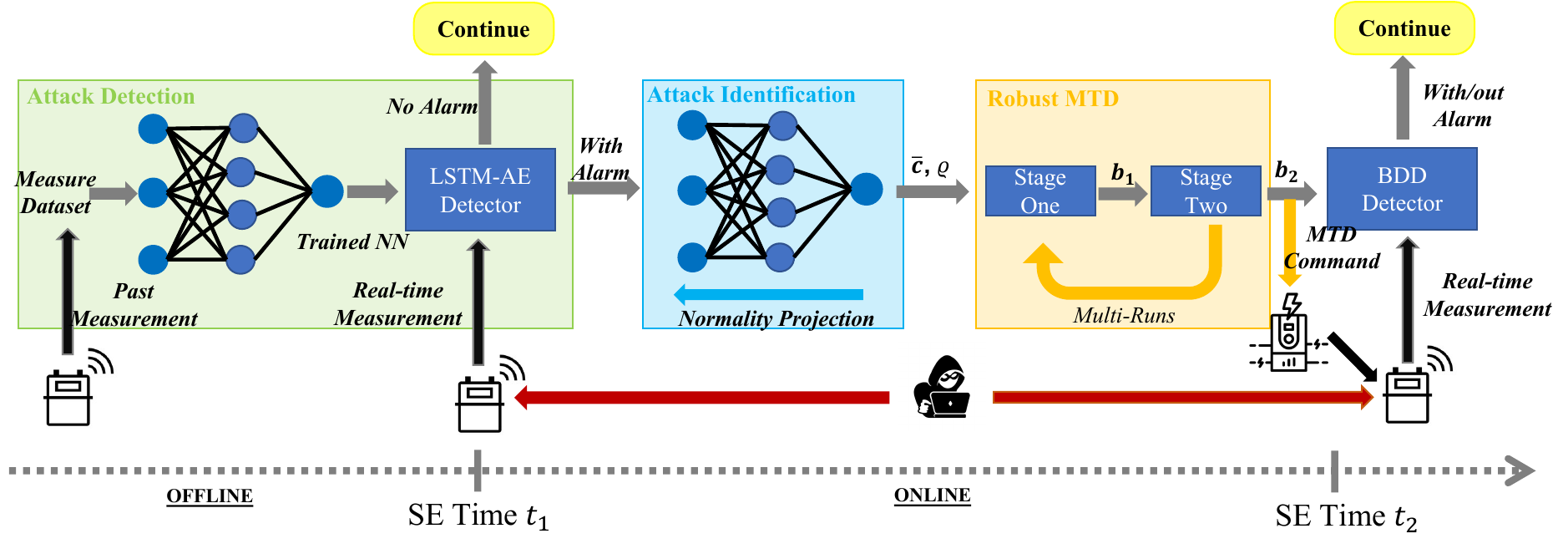}
     \caption{The DDET-MTD framework in one execution cycle.}
     \label{fig:framework}
\end{figure*}

As shown in Fig. \ref{fig:framework}, the proposed DDET-MTD has three successive components in one execution cycle. First, the LSTM-AE detector in Section \ref{para:lstm-ae} is trained on the normal dataset offline and then tests the sensor measurement collected from SCADA in real-time operation. If a positive alarm is raised at the SE time $t_1$, an attack identification algorithm is implemented to approximately extract the attack vector in the second component using the same neural network. The attack identification serves as the bridge between the data and physics by applying the extracted attack knowledge to the MTD design, in the meantime, reduces the execution cost of MTD and improves its hiddenness. 
In the last component, based on the identified attack, a robust MTD algorithm is triggered to verify the positive alarm from the LSTM-AE detector at the next SE time $t_2$. Intuitively, the false alarms from the data-driven detector can be sufficiently rejected by the subsequent MTD due to the controllable FPR of MTD.


\subsection{Physics-Informed Attack Identification}
\label{sec:identification}

The LSTM-AE detector defines a manifold for normal measurement. Therefore, the attack identification can be achieved by first recovering the normal measurement toward the manifold of the LSTM-AE detector. Following Section \ref{para:lstm-ae}, given a continuous measurement set $\mathcal{Z}_i = \{\bm{z}_1,\bm{z}_2,\dots,\bm{z}_T\}$ with positive alarm by the LSTM-AE detector, we assume that only the last measurement vector is anomalous. Let the anomalous attack and recovered measurement be $\bm{z}_a$ and $\bm{z}_T^{nor}$, respectively. To explicitly encode the measurement equation \eqref{eq:measure_nonlinear_1}, the recovered measurement can be written as
\begin{equation}\label{eq:recover_z}
    \bm{z}_T^{nor} = \bm{h}(\bm{v}^{nor}_{R,T}, \bm{v}^{nor}_{I,T})
\end{equation}
where $\bm{v}^{nor}_{R,T}$ and $\bm{v}^{nor}_{I,T}$ are the recovered real and imaginary voltage vectors. Here, the rectangular form on complex number is used to ensure stable back-propagation in Neural Network. Let $\mathcal{Z}^{nor}_i = \{\bm{z}_1,\bm{z}_2,\dots,\bm{z}_T^{nor}\}$. An energy function measuring the distance from $\mathcal{Z}^{nor}_i$ to the normality manifold defined by the LSTM-AE can be written as:
\begin{equation}\label{eq:energy_function}
    \begin{aligned}
         \mathcal{E}(\bm{v}^{nor}_{R,T}, \bm{v}^{nor}_{I,T}) = & \mathcal{L}(\mathcal{Z}_i^{nor}) + \beta_R\|\bm{v}^{nor}_{R,T}-\bm{v}^{nor}_{R,a}\|_1 \\ &  \quad  + \beta_I\|\bm{v}^{nor}_{I,T}-\bm{v}^{nor}_{I,a}\|_1
    \end{aligned}
\end{equation}

Eq. \eqref{eq:energy_function} can be viewed as a non-linear Lasso regression on decision variable $(\bm{v}^{nor}_{R,T}, \bm{v}^{nor}_{I,T})$ where the projection of the attack measurement $\bm{z}_a$ on the manifold of LSTM-AE is calculated with physical information \eqref{eq:recover_z} considered. In detail, the first term in \eqref{eq:energy_function} is the reconstruction loss \eqref{eq:lstm_loss} of the recovered normal measurement $\bm{z}_T^{nor}$, while the second and third terms penalise the difference between real and imaginary-part voltage deviations with weights $\beta_R$ and $\beta_I$, respectively. Since the attack is usually sparse, the $l_1$-norm is used to regularise the number of attacked states. The $l_1$-norm is also less sensitive to attack vector than the $l_2$-norm used in $\mathcal{L}(\mathcal{Z}_i^{nor})$ of \eqref{eq:lstm_loss}.   

\begin{figure}[h]
    \centering
    \includegraphics[width=0.78\linewidth]{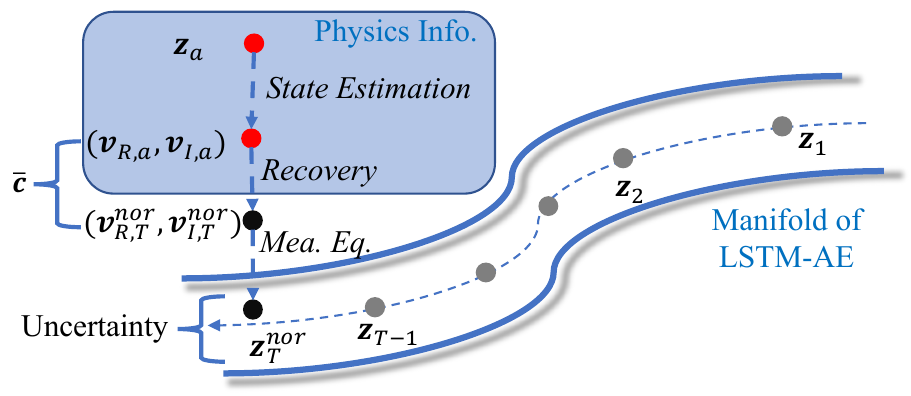}
    \caption{Illustration on attack identification algorithm.}
    \label{fig:recovery}
\end{figure}

The attack identification algorithm is illustrated in Fig. \ref{fig:recovery} and Algorithm \ref{alg:recovery}. As shown in Fig. \ref{fig:recovery}, the physics information is encoded through the measurement equation \eqref{eq:measure_nonlinear_1} and the SE \eqref{eq:se_nonlinear} when projecting the attack measurement onto the normal manifold defined by the LSTM-AE detector. Therefore, the main component of Algorithm \ref{alg:recovery} is to recover $\bm{z}_T^{nor}$ seen by both the BDD and the LSTM-AE detector. In line 3, the state estimation of the previous measurement is used as the warm start. \texttt{Adam} Optimiser \cite{kingma2014adam} is used to minimise the weighted loss $\mathcal{E}^k$ \eqref{eq:energy_function} by backpropagation with step size $lr_{\text{identifier}}$. Iteration in lines 5-14 is terminated if the reconstruction loss \eqref{eq:lstm_loss} is lower than the threshold $\tau_{\text{lstm}}$ or the maximum iteration number $ite_{\text{max}}$ is achieved. The minimum iteration number $ite_{\text{min}}$ is designed for warm-up purposes. Finally, the attack vector is identified by subtraction in line 15. 

Given the $i$-th attack in an attack index set $\mathcal{I}_a$, the attack identification uncertainty set can be empirically determined as $\mathcal{C}_i = \{\bm{c}'|\|\bm{c}'-\bar{\bm{c}}_i\|_2^2<\varrho^2\}$ where $\varrho$ is the empirical upper bound on the deviation between the identified attack vector $\bar{\bm{c}}_i$ and the ground truth $\bm{c}_i$ for any $\forall i \in \mathcal{I}_a$.

\begin{algorithm}[t]
    \footnotesize
    \SetKwInOut{Input}{Input}
    \SetKwInOut{Output}{Output}

    \Input{$\mathcal{L}(\cdot)$, $\mathcal{Z}_i$, $lr_{\text{identifier}}$, $\beta_{R}$, $\beta_I$, $\tau_{\text{lstm}}$, $ite_{\text{min}}$, $ite_{\text{max}}$}
    \Output{Identified attack vector $\bar{\bm{c}} = (\bar{\bm{c}}_R, \bar{\bm{c}}_I)$}
    
    
    Do state estimation on $\bm{z}_T^a$ as $\bm{v}_{R,T}^{a}$ and $\bm{v}_{I,T}^{a}$, and on $\bm{z}_{T-1}$ as $\bm{v}_{R,T-1}$ and $\bm{v}_{I,T-1}$
    
    $k=1$
    
    $\bm{v}_{R,T}^{k} = \bm{v}_{R,T-1}$, $\bm{v}_{I,T}^{k} = \bm{v}_{I,T-1}$ \tcc{warm start}
    
    Initialize \texttt{Adam} optimiser with $lr_{\text{identifier}}$
    
    \While{$k \leq ite_{\text{max}}$}
    {
    $\bm{z}_T^k = \bm{h}(\bm{v}_{R,T}^{k}, \bm{v}_{I,T}^{k})$
    
    $\mathcal{Z}^k = \text{combine}\{\mathcal{Z}_i[1:T-1], \bm{z}_T^k\}$
    
    $\mathcal{E}^k = \mathcal{L}(\mathcal{Z}^k_i) + \beta_R \cdot \|\bm{v}_{R,T}^{k} - \bm{v}_{R,T}^{a}\|_1 + \beta_I \cdot \|\bm{v}_{I,T}^{k} - \bm{v}_{I,T}^{a}\|_1$
    
    \If{$k\geq ite_{\text{min}}$ or $\mathcal{L}(\mathcal{Z}^k_i) < \tau_{\text{lstm}}$}
    {break}
    
    $(\bm{v}_{R,T}^{k+1}, \bm{v}_{I,T}^{k+1}) \xleftarrow{\texttt{Adam}} \arg\min{\mathcal{E}^k}$
    
    $k\leftarrow k+1$
    }
    
    
    $\bar{\bm{c}}_R = \bm{v}_{R,T}^a - \bm{v}_{R,T}^{k}$, $\bar{\bm{c}}_I = \bm{v}_{I,T}^a - \bm{v}_{I,T}^{k}$
    
    \caption{Attack Identification}
    \label{alg:recovery}
\end{algorithm}


To sum up, Algorithm \ref{alg:recovery} guarantees that the recovered measurement can bypass the LSTM-AE detector and BDD. Therefore, the recovered state obeys the physics rules of power system. It also takes advantage of the formulation on FDI attacks so that the identified attack vector can lead to a stealthy attack, which further improves the identification accuracy. 

\subsection{Hidden and Effective MTD Algorithm}

In the third component of DDET-MTD, the positive alarm from the LSTM-AE detector and the identified state attack vector can be used to trigger and design the MTD algorithm. Before introducing the idea of event triggering, the MTD algorithm is formulated as follows. 
\begin{subequations}\label{eq:mtd_abstract}
    \begin{alignat}{2}
       \min_{\bm{b}'\in\mathcal{B}} & \quad \text{$\mathbb{P}$(Attacker can detect the MTD)} \\
       \text{subject to} & \quad \text{$\mathbb{P}$(Operator can detect the attack) $\geq \rho$}  \label{eq:mtd_abstract_con}
    \end{alignat}
\end{subequations}

Although the hiddenness of MTD is essential to deceive the prudent attacker, we argue that the main target of MTD is to detect the ongoing attack with high detection rate. Therefore, \eqref{eq:mtd_abstract} is designed to minimize the attacker's chances to notice the existence of MTD, subject to a specific detection accuracy $\rho$ on the attack. However, this optimisation problem is intrinsically hard to solve for two reasons. First, both the cost and the constraint in \eqref{eq:mtd_abstract} are probabilistic and nonconvex, so the convergence property and global optimality are difficult to guarantee. Second, to guarantee the detection accuracy, it requires the exact knowledge of the attack vector, which cannot be known in advance. To address the first problem, local linearisations are introduced in the measurement equation \eqref{eq:measure_nonlinear_1} on which convex relaxation is applied. For the second, a robust two-stage optimisation problem is established based on the set of identification uncertainty set $\mathcal{C}_i$ from Algorithm \ref{alg:recovery}.

\subsubsection{Approximations of MTD Hiddenness}

By explicitly considering the influence of susceptance on the measurement, the measurement equation \eqref{eq:measure_nonlinear_1} can be rewritten as $\bm{z}=\bm{h}(\bm{v},\bm{b}) + \bm{e}$. Under normal operation (no attack and MTD), the last iteration of SE is to solve the following normal equation:
\begin{equation}\label{eq:measure_linear_no_mtd}
    {\bm{z}-\bm{h}(\widehat{\bm{v}},\bm{b}_0)} = \bm{H}_{\widehat{\bm{v}}}({\bm{v}-\widehat{\bm{v}}})+\bm{e}
\end{equation}
where $\bm{H}_{\widehat{\bm{v}}} = \left[\frac{\partial \bm{h}(\bm{v},\bm{b}_0)}{\partial \bm{v}}\right]_{\bm{v}=\widehat{\bm{v}}}$ and $\widehat{\bm{v}}$ is the state estimated before MTD. For the above system, the residual can be derived
\begin{equation}\label{eq:residual_no_mtd}
    \gamma(\bm{z},\bm{b}_0)=\|\bm{R}^{-\frac{1}{2}}\bm{S}_{\widehat{\bm{v}}}\bm{e}\|_2^2
\end{equation}
where $\bm{S}_{\widehat{\bm{v}}} = \mathcal{S}_{\bm{R}^{-1}}(\bm{H}_{\widehat{\bm{v}}})$. $\gamma(\bm{z},\bm{b}_0)$ also follows the $\chi^2$ distribution with DoF $P-2N$.

When the MTD is triggered, both $\bm{b}$ and $\bm{v}$ will be deviated from the stationary point. The first-order Taylor expansion around $(\widehat{\bm{v}},\bm{b}_0)$ is written as:
\begin{equation}\label{eq:measure_linear_with_mtd}
    {\bm{z}'-\bm{h}(\widehat{\bm{v}},\bm{b}_0)}=\bm{H}_{\widehat{\bm{v}}}(\bm{v}'-\widehat{\bm{v}}) + \bm{H}_{\bm{b}_0}(\bm{b}'-\bm{b}_0)+\bm{e}
\end{equation}
where $\bm{H}_{\bm{b}_0} = \left[\frac{\partial \bm{h}(\widehat{\bm{v}},\bm{b})}{\partial \bm{b}_0}\right]_{\bm{b}=\bm{b}_0}$.

Combining \eqref{eq:measure_linear_no_mtd}-\eqref{eq:measure_linear_with_mtd}, the attacker's residual on the post-MTD measurement $\Delta \bm{z}'$ becomes:
\begin{equation*}
    \begin{aligned}
        \gamma(\bm{z}',\bm{b}_0) & =\|\bm{R}^{-\frac{1}{2}}\bm{S}_{\widehat{\bm{v}}}(\bm{H}_{\widehat{\bm{v}}}(\bm{v}'-\widehat{\bm{v}}) + \bm{H}_{\bm{b}_0}(\bm{b}' - \bm{b}_0) + \bm{e})\|_2^2 \\
        & = \|\bm{R}^{-\frac{1}{2}}\bm{S}_{\widehat{\bm{v}}}( \bm{H}_{\bm{b}_0}(\bm{b}'-\bm{b}_0) + \bm{e})\|_2^2
    \end{aligned}
\end{equation*}
in which the second equality is due to the fact that $\bm{S}_{\hat{\bm{v}}} \bm{H}_{\hat{\bm{v}}} = 0$. Meanwhile, $\gamma(\bm{z}',\bm{b}_0)$ follows the non-central $\chi^2$ distribution (NCX) with non-centrality parameter:
\begin{equation}\label{eq:appro_hid}
    \lambda(\bm{z}',\bm{b}_0) = \|\bm{R}^{-\frac{1}{2}}\bm{S}_{\widehat{\bm{v}}} \bm{H}_{\bm{b}_0}(\bm{b}'-\bm{b}_0)\|_2^2
\end{equation}

Since the probability that the MTD is detected by the attacker increases monotonically as $\lambda(\bm{z}',\bm{b}_0)$ increases \cite{xu2022robust}, $\lambda(\bm{z}',\bm{b}_0)$ should be minimised. This result is coherent to the findings in \cite{liu2021optimal, tian2018enhanced, liu2022explicit} where the measurement change before and after MTD should be small. Note that both $\bm{S}_{\widehat{\bm{v}}}$ and $\bm{H}_{\bm{b}_0}$ are constants for a given load condition. Meanwhile, $\bm{H}_{\bm{b}_0}$ can be derived analytically using similar methods in \cite{liu2021interior}.

\subsubsection{Approximation of MTD Effectiveness}

To accelerate the convergence speed and performance of SE, dishonest SE is widely used, in which the Jacobian matrix remains unchanged throughout the iteration \cite{zhang2019analysis}. The last iteration of dishonest SE on $\bm{z}'$ is represented as:
\begin{equation}\label{eq:mtd_eff_linear}
    {\bm{z}' - \bm{h}'(\widehat{\bm{v}}')}=\bm{H}_{\bm{v}_0}'({\bm{v}'-\widehat{\bm{v}}'})+\bm{e}
\end{equation}
where $\widehat{\bm{v}}'$ is the estimated state of $\bm{z}'$ and $\bm{H}_{\bm{v}_0}' = \left[\frac{\partial \bm{h}'(\bm{v})}{\partial \bm{v}}\right]_{\bm{v}=\bm{v}_0}$. 

The residual of the above system is derived as $\gamma(\bm{z}',\bm{b}') = \|\bm{R}^{-\frac{1}{2}}\bm{S}_{\bm{v}_0}'\bm{e}\|_2^2$ where $\bm{S}'_{\bm{v}_0} = \mathcal{S}_{\bm{R}^{-1}}(\bm{H}'_{\bm{v}_0})$. Similarly, $\gamma(\bm{z}',\bm{b}')$ follows the $\chi^2$ distribution with DoF $P-2N$. 

When an attack exists, $\bm{a}=\bm{h}(\widehat{\bm{v}}'_a + \bm{c})-\bm{h}(\widehat{\bm{v}}'_a)$ where $\widehat{\bm{v}}_a'$ is the estimated state from the attacker after the MTD is triggered. As required by the MTD hiddenness, the difference in pre- and post-MTD measurements is minimised. Therefore, it is reasonable to assume that $\widehat{\bm{v}}_a'$ is close to $\widehat{\bm{v}}$. Following Assumption Two, for small state injection, the attack vector can be approximated as $\bm{a} = \bm{H}_{\widehat{\bm{v}}}\bm{c}$ \cite{liu2019joint}. Consequently, the non-centrality parameter of the post-MTD measurement under attack is approximated as:
\begin{equation}\label{eq:appro_eff}
    \lambda(\bm{z}_a',\bm{b}')=\|\bm{R}^{-\frac{1}{2}}\bm{S}_{\bm{v}_0}'\bm{H}_{\widehat{\bm{v}}}\bm{c}\|_2^2
\end{equation}

\subsubsection{Attack-Aware Robust MTD Reformulation}

Based on the approximations of the hiddenness \eqref{eq:appro_hid} and effectiveness \eqref{eq:appro_eff} of MTD, the probabilistic optimisation problem \eqref{eq:mtd_abstract} becomes nonprobabilistic for a given attack $\bm{c}$:
\begin{subequations}\label{eq:mtd_1}
    \begin{alignat}{2}
       \min_{\bm{b}'\in\mathcal{B}} & \quad \lambda(\bm{z}',\bm{b}_0) \label{eq:mtd_1_obj}  \\
       \text{s.t.} & \quad \lambda(\bm{z}_a',\bm{b}') \geq \lambda_c(\rho)  \label{eq:mtd_1_con}
    \end{alignat}
\end{subequations}

In \eqref{eq:mtd_1_obj}, only when $\lambda(\bm{z}',\bm{b}_0) = 0$, the MTD can be 100\% hidden to the attacker. In most of the cases, the MTD hiddenness and effectiveness are proved to be contradictory \cite{tian2018enhanced, zhang2020hiddenness, liu2021optimal}. In \eqref{eq:mtd_1_con}, the probability constraint \eqref{eq:mtd_abstract_con} is converted non-probabilistic. In fact, there is a $\lambda_c(\rho)$ such that the detection rate at $\bm{c}$ is equal to $\rho$ \cite{xu2022robust}:
\begin{equation}\label{eq:ncx_1}
    \mathbb{P}(\gamma\geq\tau(\alpha)) = \int_{\tau(\alpha)}^{\infty}f_\chi(\gamma|\kappa,\lambda_c)=\rho
\end{equation}
where $f_\chi(\gamma|\kappa,\lambda_c)$ represents the density function of the NCX distribution with DoF $\kappa=P-2N$ and the non-centrality parameter equals $\lambda_c$.

The optimisation \eqref{eq:mtd_1} still requires exact knowledge of the attack vector $\bm{c}$, which is not available for the operator. Therefore, a robust reformulation of \eqref{eq:mtd_1} is derived by guaranteeing the lowest detection rate for the attacks in the attack uncertainty set $\mathcal{C}$ defined in Section \ref{sec:identification}: 
\begin{subequations}\label{eq:mtd_bilevel}
    \begin{alignat}{2}
       \min_{\bm{b}'}    & \quad \lambda(\bm{z}',\bm{b}_0)  \\
       \text{s.t.} & \quad \bm{b}'\in\mathcal{B} \\ 
                         & \quad \min_{\bm{c}'\in\mathcal{C}} \lambda(\bm{z}_a',\bm{b}') \geq \lambda_c(\rho)  \label{eq:mtd_bilevel_nest}
    \end{alignat}
\end{subequations}

Problem \eqref{eq:mtd_bilevel} is a bilevel optimization problem \cite{sinha2017review}. The objective of the upper level is to decrease the chance that the attacker detects MTD. The decision variable in upper level is the MTD setpoint $\bm{b}'$ and the constraint on $\bm{b}'$ is the permissible set of D-FACTS devices $\mathcal{B}$. At the lower level, the objective function is to find the state injection that results in the lowest detection rate, subject to the set of uncertainties $\mathcal{C}$. Note that the upper level decision variable $\bm{b}'$ is nested at the lower level parametrically. The nesting structure robustly ensures that all possible attacks in $\mathcal{C}$ can be detected with predefined probability $\rho$. 

To simplify the analysis, only active power flow measurements are considered for MTD effectiveness, as active power is more important in state estimation and sensitive to changes in the voltage phase angle\cite{liu2021optimalcoding}. 
As a result, the Jacobian matrix in \eqref{eq:mtd_eff_linear} can be analytically written as:
\begin{equation}\label{eq:jacobian}
    \bm{H}_{\bm{v}_0} = \underbrace{\bm{V}\cdot\bm{G}\cdot\bm{A}_r^{s}}_{\bm{C}}-\bm{V}\cdot\bm{B}\cdot  \bm{A}_r^{c}
\end{equation}
where $\bm{V} = [(\bm{C}_f\bm{v}_0)\circ(\bm{C}_t\bm{v}_0)]$; $\bm{G} = [\bm{g}]$; $\bm{B} = [\bm{b}]$; $\bm{A}_{r}^{s} = [\sin{\bm{A\theta_0}}]\bm{A}_r$; and $\bm{A}_{r}^{c} = [1/\bm{\bm{t}}][\cos{\bm{A\theta_0}}]\bm{A}_r$. $\bm{A}_r$ is the reduced incidence matrix by removing the column that represents the reference bus from the incidence matrix $\bm{A}$; $\bm{t}$ is the vector of the transformer tap ratio. The detection threshold corresponding to the active power flow measurements is denoted as $\lambda_c'$. Intuitively, guaranteeing the detection rate on a subset of the measurement can also guarantee the detection rate on the full measurement due to the increased redundancy. 

As proved by \cite{xu2022robust}, only when the attack strength is greater than a certain threshold can $\lambda(\bm{z}_a',\bm{b}') \geq \lambda_c'$ be satisfied. Therefore, despite the non-linearity and non-convexity, \eqref{eq:mtd_bilevel} may not have a feasible solution. As a result, \eqref{eq:mtd_bilevel} is separated into two stages. In stage one, the feasibility of constraint \eqref{eq:mtd_bilevel_nest} is checked by maximizing its left hand side. The optimal solution of stage one is then used as the feasible warm start in stage two to improve its hiddenness.

\subsubsection{Convex Stage-One Optimisation} 

In stage one, the feasibility of constraint \eqref{eq:mtd_bilevel_nest} is checked by maximizing the detection rate on the worst-case attack in $\mathcal{C}$
\begin{equation}\label{eq:mtd_stage_one}
       \max_{\bm{b}'\in\mathcal{B}} \min_{\bm{c}'\in\mathcal{C}}\lambda(\bm{z}_a',\bm{b}')
\end{equation}

Multi-run strategy is required to solve the non-convex problem \eqref{eq:mtd_stage_one} with different starting points in $\mathcal{B}$. For each run, an equivalent convex reformulation is derived as follows:

\begin{proposition}\label{theorem:mtd_stage_one}
\normalfont
Define auxiliary variable $\omega\in\mathbb{R}$, $\nu\in\mathbb{R}$, $\bm{H}_1 = \bm{R}^{-\frac{1}{2}}\bm{H}_{\widehat{\bm{v}}}$, and $\bm{H}_0' = \bm{R}^{-\frac{1}{2}}\bm{H}_{\bm{v}_0}'$. The problem \eqref{eq:mtd_stage_one} is equivalent to the following:
\begin{subequations}\label{eq:mtd_stage_one_lmi}
    \begin{alignat}{2}
    \max_{\bm{b}',\nu,\omega} & \quad \omega \\
    \text{s.t.} 
    & \quad [\bm{b}'] - [\bm{b}^-] \succeq 0,  [\bm{b}^+] - [\bm{b}'] \succeq 0 \label{eq:mtd_stage_one_dfact} \\
    & \quad \nu \geq 0 \label{eq:mtd_stage_one_dual_variable}\\
    & \begin{bmatrix}
        \nu(\bar{\bm{c}}^T\bar{\bm{c}} - \varrho^2) - \omega & \nu\bar{\bm{c}}^T & \bm{O} \\
        \star & \nu\bm{I}+\bm{H}_1^T\bm{H}_1 & \bm{H}_1^T\bm{H}_0' \\
        \star & \star & {\bm{H}_0'}^{T}\bm{H}_0' \label{eq:mtd_stage_one_lmi_matrix}
    \end{bmatrix} \succeq 0 
    \end{alignat}
\end{subequations}
\end{proposition}

\begin{proof}
The proof can be found in Appendix \ref{sec:app_1}.
\end{proof}

Referring to \eqref{eq:jacobian}, the only nonlinearity of \eqref{eq:mtd_stage_one_lmi} is in the last block-diagonal entry of \eqref{eq:mtd_stage_one_lmi_matrix}. To linearise ${\bm{H}_0'}^T\bm{H}_0'$, iterative algorithm is designed with starting point $\bm{b}_0$ and the following proposition is derived:

\begin{proposition}\label{theorem:mtd_stage_one_ite}
Let $\bm{C}^N=\bm{R}^{-\frac{1}{2}}\bm{C}$ and $\bm{V}^N=\bm{R}^{-\frac{1}{2}}\bm{V}$. Define $\bm{b}_k$ as the feasible solution of the $k$-th iteration. A sufficient convex condition for \eqref{eq:mtd_stage_one_lmi_matrix} is
\begin{equation}\label{eq:mtd_stage_one_lmi_update}
    \begin{bmatrix}
        \nu(\bar{\bm{c}}^T\bar{\bm{c}} - \varrho^2) - \omega & \nu\bar{\bm{c}}^T & \bm{O} \\
        \star & \nu\bm{I}+\bm{H}_1^T\bm{H}_1 & \bm{H}_1^T\bm{H}_0' \\
        \star & \star & \bm{H}_{\text{update}}
    \end{bmatrix} \succeq 0
\end{equation}
with $\bm{H}_{\text{update}} = (\bm{V}^N[\bm{b}_k]\bm{A}_r^c)^T(\bm{C}^N+\bm{V}^N[\bm{b}']\bm{A}_r^c) + (\bm{C}^N+\bm{V}^N[\bm{b}']\bm{A}_r^c)^T(\bm{V}^N[\bm{b}_k]\bm{A}_r^c) - (\bm{V}^N[\bm{b}_k]\bm{A}_r^c)^T(\bm{V}^N[\bm{b}_k]\bm{A}_r^c)$.
\end{proposition}

\begin{proof}
    The proof can be found in Appendix \ref{sec:app_ite}.
\end{proof}

In summary, at the $k$-th iteration, the following convex programming is solved until convergence, though it may not converge to the global optimality of \eqref{eq:mtd_stage_one} and \eqref{eq:mtd_stage_one_lmi}.
\begin{equation}\label{eq:mtd_stage_one_iteration}
    \begin{aligned}
    \max_{\bm{b}',\mu,\omega} & \quad \omega \\
    \text{s.t.} 
    & \quad \eqref{eq:mtd_stage_one_dfact}, \eqref{eq:mtd_stage_one_dual_variable}, \eqref{eq:mtd_stage_one_lmi_update}
    \end{aligned}
\end{equation}



\subsubsection{Convex Stage-Two Optimisation}


The stage-one problem checks the feasibility of \eqref{eq:mtd_bilevel_nest}. In detail, if the optimal solution $\omega^*$ of \eqref{eq:mtd_stage_one_lmi} (or similarly the final iteration of \eqref{eq:mtd_stage_one_iteration}) is greater than $\lambda_c'$, the original bilevel problem \eqref{eq:mtd_bilevel} can be solved with the optimal point of \eqref{eq:mtd_stage_one_lmi} as a feasible warm start. Otherwise, the threshold in \eqref{eq:mtd_bilevel_nest} should be reduced to $\omega^\star$ to have a feasible solution. In either situation, denoting the threshold of constraint \eqref{eq:mtd_bilevel_nest} after stage one as $\omega$, the following proposition gives a feasible and convex reformulation to \eqref{eq:mtd_bilevel} in which the MTD effectiveness is guaranteed to the level determined by stage one while the hiddenness is improved.

\begin{proposition}\label{theorem:mtd_stage_two}
\normalfont
With all variables and parameters defined as in Proposition \ref{theorem:mtd_stage_one}, and let auxiliary variable $\phi \geq 0$, $\bm{H}_{\text{hid}} = \bm{R}^{-\frac{1}{2}}\bm{S}_{\widehat{\bm{v}}}\bm{H}_b$. The bilevel optimisation problem \eqref{eq:mtd_bilevel} with $\lambda_c(\rho)$ replaced by $\omega$ can be solved by
\begin{subequations}\label{eq:mtd_stage_two_lmi}
    \begin{alignat}{2}
    \min_{\bm{b}',\nu,\phi} & \quad \phi \\
    \text{s.t.} 
    & \quad \eqref{eq:mtd_stage_one_dfact}, \eqref{eq:mtd_stage_one_dual_variable}, \eqref{eq:mtd_stage_one_lmi_matrix}\\
    & \quad \begin{bmatrix}
        \phi & (\bm{b}'-\bm{b}_0)^T\bm{H}^T_{\text{hid}} \\
        \star & \bm{I}
    \end{bmatrix} \succeq 0 \label{eq:mtd_stage_two_new}
    \end{alignat}
\end{subequations}

\end{proposition}

\begin{proof}
The proof can be found in Appendix \ref{sec:app_2}.
\end{proof}

Similarly, the non-convexity in ${\bm{H}_0'}^{T}\bm{H}_0'$ can be solved iteratively by the sufficient condition described in Proposition \ref{theorem:mtd_stage_one_ite}. This results in an iterative algorithm to solve the stage-two problem:

\begin{equation}\label{eq:mtd_stage_two_iteration}
    \begin{aligned}
    \max_{\bm{b}',\mu,\phi} & \quad \phi \\
    \text{s.t.} 
    & \quad \eqref{eq:mtd_stage_one_dfact}, \eqref{eq:mtd_stage_one_dual_variable}, \eqref{eq:mtd_stage_one_lmi_update}, \eqref{eq:mtd_stage_two_new}
    \end{aligned}
\end{equation}
where $\omega$s in \eqref{eq:mtd_stage_two_lmi} and \eqref{eq:mtd_stage_two_iteration} are constants determined by the optimum of stage one. 

To conclude, the two-stage optimisation in DDET-MTD has been developed to guarantee the effectiveness of MTD while improving hiddenness. Based on convex relaxation, the hidden and effective MTD can be designed as follows:
\begin{enumerate}
    \item Solve the stage-one problem \eqref{eq:mtd_stage_one_iteration} iteratively with different start point $\bm{b}_0$. Store the multi-run results of $\bm{b}'$ in a set $\mathcal{D}^{\text{one}}$ and the corresponding cost $\omega$ into a set $\mathcal{G}^{\text{one}}$.
    \item If the largest cost in $\mathcal{G}^{\text{one}}$ is smaller than $\lambda_c'$, use the corresponding susceptance in $\mathcal{D}^{\text{one}}$ as a warm start in stage-two problem \eqref{eq:mtd_stage_two_iteration} and solve it iteratively.
    \item If the largest cost in $\mathcal{G}^{\text{one}}$ is larger than or equal to $\lambda_c'$, define the index set $\mathcal{I}^{\text{two}} = \{i|\omega_i \geq \lambda_c', \omega_i\in\mathcal{G}^{\text{one}}\}$ and candidate warm-start susceptance set $\mathcal{D}^{\text{two}} = \{\mathcal{D}[i], i\in\mathcal{I}^{\text{two}}\}$. For each $\bm{b}\in\mathcal{D}^{\text{two}}$, iteratively solve stage-two problem \eqref{eq:mtd_stage_two_iteration}. The optimal susceptance is returned with the smallest cost.
    
\end{enumerate}

The detailed algorithm can be found in Appendix \ref{sec:app_algorithm}.

\color{black}

\section{Simulations and Results}\label{sec:simulation}


\subsection{Simulation Settings}\label{sec:sim_settings}

\subsubsection{Model Configurations}

The proposed DDET-MTD algorithm is tested on the IEEE case-14 system \cite{zimmerman2018matpower}. Although we have derived the theoretical analysis using simplified models, all the simulations are implemented under full AC condition. Real-time load consumptions and photovoltaic generations are assigned to each bus for four months using a similar method in \cite{wang2021multi}. The load data is interpolated to 5-min resolution, resulting in over 35k data in total. For each operation instance, AC-OPF is solved by \texttt{PyPower} \cite{zimmerman2018matpower}. The standard deviation of the measurement noise is set to 2\% of the default values in the case-14 system case file. The FPR of BDD is set as $\alpha = 2\%$. The MTD threshold $\lambda_c$ and $\lambda_c'$ are determined by \eqref{eq:ncx_1} with $\rho=1-\alpha = 98\%$. LSTM-AE attack detection and identification algorithms are trained and implemented using \texttt{PyTorch}\cite{paszke2019pytorch} with hyperparameters summarised in Table \ref{tab:ddd_para}. The data set is separated into 60\% training, 20\% validation, and 20\% test sets. Throughout the simulation, random sparse AC-FDI attacks are generated with the number of attacked buses equal to 1-3, and the strength of the attacks is set as $\pm10\%-20\%$ and $\pm20\%-30\%$ of the normal state solved from the real-time measurements. For example, the pair $(2,0.3)$ means that there are two buses being attacked with strength at random in $\pm 0.2-0.3$. In the simulation, 200 attacks are randomly generated from the entire test set for each type of attack. Without losing generality, all the branches are equipped with D-FACTS devices and the maximum reactance perturbation ratio is 50\% \cite{lakshminarayana2020cost}. In addition, the convex MTD optimisation problems are solved by \texttt{CVXPY}\cite{diamond2016cvxpy} with \texttt{MOSEK} solver. Hyperparameters for stage-one and stage-two optimisations are summarised in Table \ref{tab:mtd_para}. 

\begin{table}[h]
    \caption{Hyperparameters for the Detector and Identifier.}
    \centering
    \footnotesize
    \begin{tabular}{c|c|c|c}\hline\hline
        \textbf{Sample Length} & 6 & \textbf{Encoder Size} & 68-48-29-10 \\\hline
        \textbf{Epochs No.} & 1000 & \textbf{Batch Size} & 32 \\\hline
        $\bm{lr_{\textbf{detector}}}$ & 0.001 & $\bm{lr_{\textbf{identifier}}}$ & 0.005 \\\hline
        \textbf{Early Stop Patience} & 10 & \textbf{Early Stop Diff.} & 0 \\\hline
        $\bm{\beta_{R}}, \bm{\beta_{I}}$ & 0.1 & \textbf{Optimizer} & Adam \\\hline
        $\bm{ite_{\textbf{max}}}$ & 1000 & $\bm{ite_{\textbf{min}}}$ & 50\\\hline\hline
    \end{tabular}
    \label{tab:ddd_para}
\end{table}

\begin{table}[h]
    \centering
    \footnotesize
    \caption{Hyperparameters for Stage-One/Two Optimisations.}
    \begin{tabular}{c|c}\hline\hline
        \textbf{Multi-Run No.} ${no}$ & 15 \\\hline \textbf{Max. iteration No.} $ite_{\text{one}}, ite_{\text{two}}$ & 100 \\\hline
        \textbf{Tolerance of stage one} ${tol}_{\text{one}}$ & 0.1 \\\hline \textbf{Tolerance of stage two} ${tol}_{\text{two}}$ & 1 \\\hline\hline
    \end{tabular}
    \label{tab:mtd_para}
\end{table}

\subsubsection{Baseline Algorithms}

Two algorithms, namely, the modified Max-Rank MTD \cite{liu2018reactance, liu2020optimal, zhang2019analysis} and (incomplete) Robust MTD \cite{xu2022robust}, are implemented for benchmarking the proposed algorithm. In Max-Rank MTD, the D-FACTS devices are perturbed within $\mu_{\text{min}}\bm{x}_i \leq |\Delta \bm{x}_i|\leq \mu_{\text{max}}\bm{x}_i$ (with $\mu_{\text{max}} > \mu_{\text{min}} > 0$) so that the rank of the composite matrix is maximised, which results in maximum detection rate under noiseless assumption. Due to the randomness of this algorithm, we simulate 1000 attacks for each attack scenario under different load conditions and record the average performance. The Robust MTD algorithm considers maximising the detection rate on the worst-case attack without any prior knowledge on the attack. Therefore, it can be viewed as a conservative formulation on DDET-MTD with an attack uncertainty set $\mathcal{C} = \mathbb{R}^{N}$. Although both baseline algorithms are periodic with SE, we also simulate their event-triggering variants. 

\subsubsection{Metrics}

Four metrics are considered throughout the simulation. 

From an attack defence perspective, Attack Detection Probability (ADP) and Defence Hiddenness Probability (DHP) can be used to evaluate the effectiveness and hiddenness of MTD as follows \cite{liu2021optimal}.
\begin{subequations}
    \begin{equation}
        \text{ADP} = \frac{\text{Number of attacks being detected}}{\text{Total number of attacks}}
    \end{equation}
    \begin{equation}
        \text{DHP} = \frac{\text{Number of MTDs not being detected}}{\text{Total number of MTDs}}
    \end{equation}
\end{subequations}

From an economical perspective, the average cost increase and the average reactance perturbation ratio due to the trigger of MTD are considered.

\subsection{LSTM-AE Detector}\label{sec:sim_ddd}

\begin{figure}
     \centering
     \begin{subfigure}[b]{0.49\linewidth}
         \centering
         \includegraphics[width=1.00\linewidth]{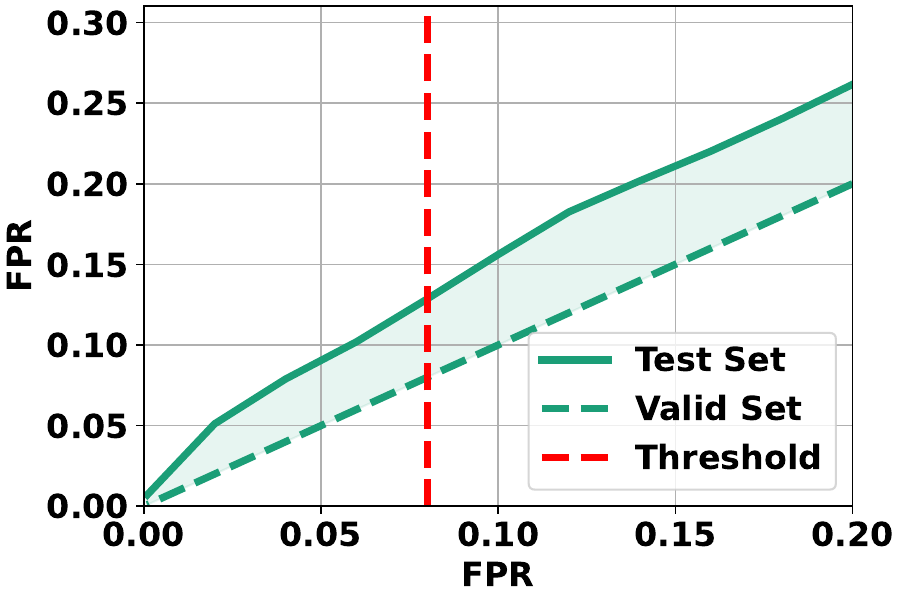}
         \caption{}
     \end{subfigure}
     \hfill
     \begin{subfigure}[b]{0.49\linewidth}
         \centering
         \includegraphics[width=\linewidth]{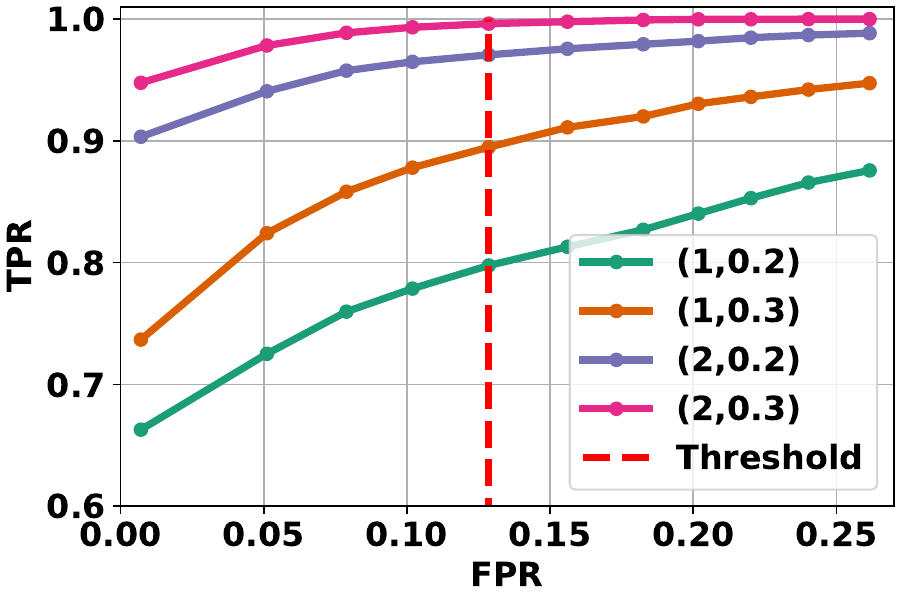}
         \caption{}
     \end{subfigure}
        \caption{Performance of LSTM-AE attack detector: (a). Generalization error (FPR) on the test dataset; (b). ROC curves on different attacks.}
        \label{fig:roc}
\end{figure}

Fig. \ref{fig:roc} illustrates the TPR and FPR of the LSTM-AE detector. Various detection thresholds $\tau_{\text{lstm}}$s are determined by the distribution of reconstruction losses in the validation set \cite{xu2020deep}. As shown in Fig. \ref{fig:roc}(a), the same detection threshold based on the FPR in the validation set can result in a higher FPR in the test set due to unseen load and PV patterns. The ROC curves on different types of attack are also summarised in Fig. \ref{fig:roc}(b), which clearly shows the trade-off between TPR and FPR. In detail, larger attack results in higher detection rate and to have 90\% TPRs on all types of attack, at least 25\% normal operation samples are incorrectly classified as attack. Since attack is rare in real-time operation, this high FPR can significantly influence the normal operation. In the following simulation, $\tau_{\text{lstm}}$ corresponding to 8.0\% FPR in the validation set is used as the detection threshold in the LSTM-AE detector, resulting in 12.84\% FPR in the test set (highlighted by the red dotted line in Fig. \ref{fig:roc}).

\subsection{LSTM-AE Identifier} \label{sec:sim_identifier}

\begin{figure}
     \centering
     \begin{subfigure}[b]{0.49\linewidth}
         \centering
         \includegraphics[width=1.00\linewidth]{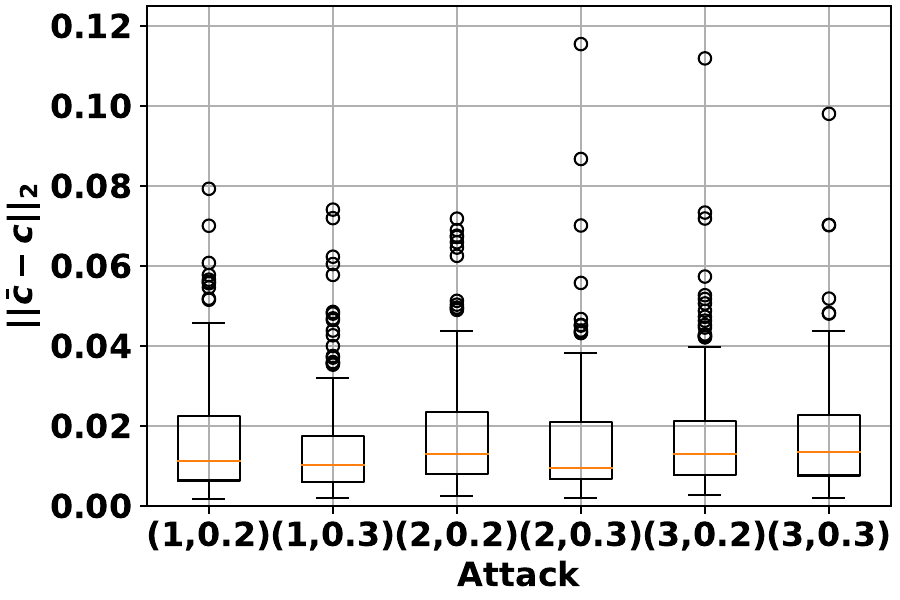}
         \caption{}
     \end{subfigure}
     \hfill
     \begin{subfigure}[b]{0.49\linewidth}
         \centering
         \includegraphics[width=\linewidth]{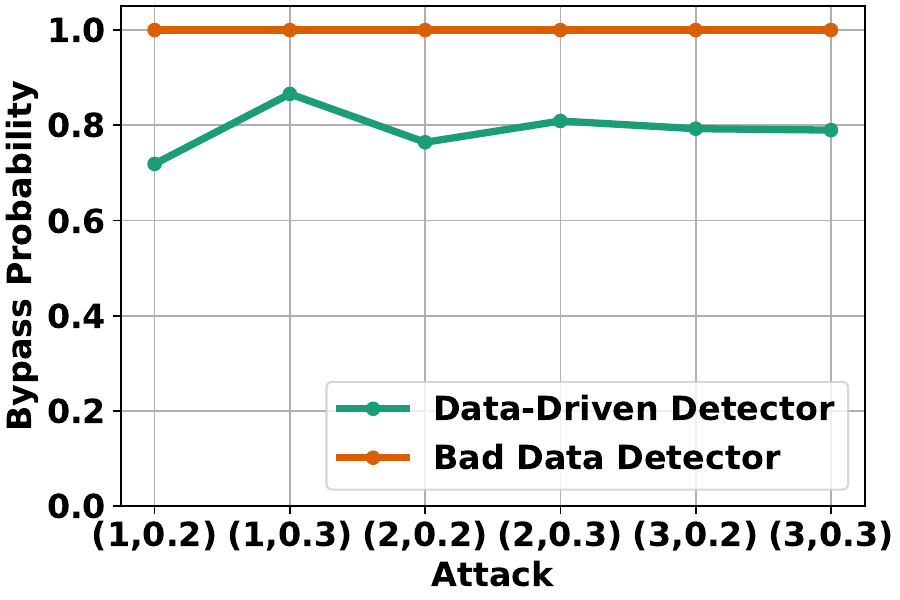}
         \caption{}
     \end{subfigure}
        \caption{Performance of LSTM-AE attack identification: (a). Identification deviation (in p.u.); (b). Probability of bypassing detectors.}
        \label{fig:iden}
\end{figure}

Fig. \ref{fig:iden} summarised the performance of the attack identification algorithm. As shown in Fig. \ref{fig:iden}(a), the average identification deviation is around 0.01p.u. and most of them are smaller than 0.04p.u.. As the average normal state angle in the simulation is 0.71p.u., the identification algorithm is accurate and stable under different attack scenarios. Fig. \ref{fig:iden}(b) tests whether the recovered measurement can bypass the BDD and LSTM-AE detector. First, since the identification algorithm filters the measurement noise by \eqref{eq:recover_z}, the recovered measurement can certainly bypass the BDD. Second, due to the existence of regularization in the energy function \eqref{eq:energy_function} and the limit of iteration numbers, only 80\% of the recovered measurement can bypass the LSTM-AE detector. Nonetheless, the reconstruction losses are much smaller than those of the attacked measurement, meaning that the recovered measurements are close to the normality manifold defined by the LSTM-AE detector. Therefore, the identified attack vector is quite accurate and can be used to guide the hidden and effective MTD algorithm.

\subsection{Properties of DDET-MTD}

In this section, we investigate the performance of the proposed DDET-MTD algorithm.

\subsubsection{Sensitivity of $\varrho$}

\begin{figure}
     \centering
     \begin{subfigure}[b]{0.49\linewidth}
         \centering
         \includegraphics[width=1.00\linewidth]{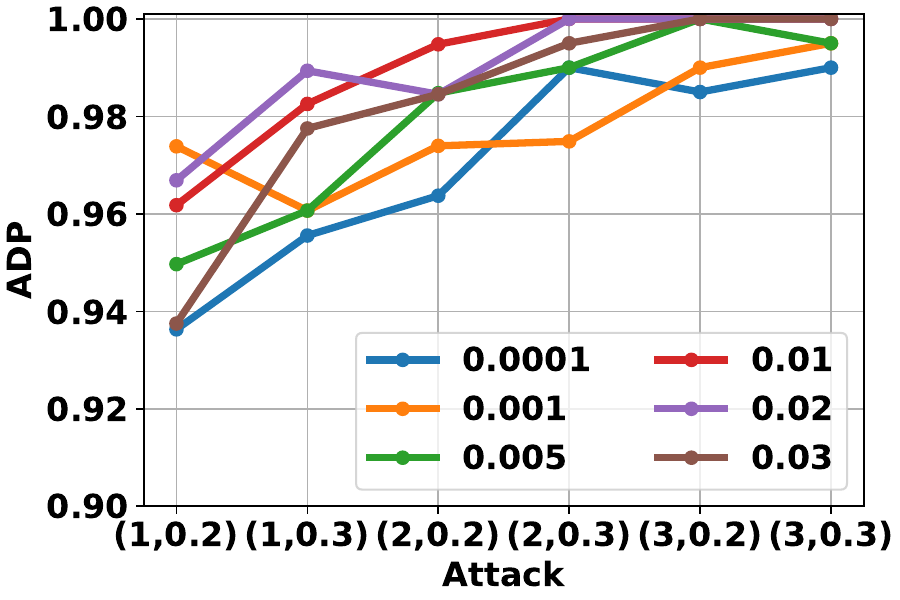}
         \caption{}
     \end{subfigure}
     \hfill
     \begin{subfigure}[b]{0.49\linewidth}
         \centering
         \includegraphics[width=\linewidth]{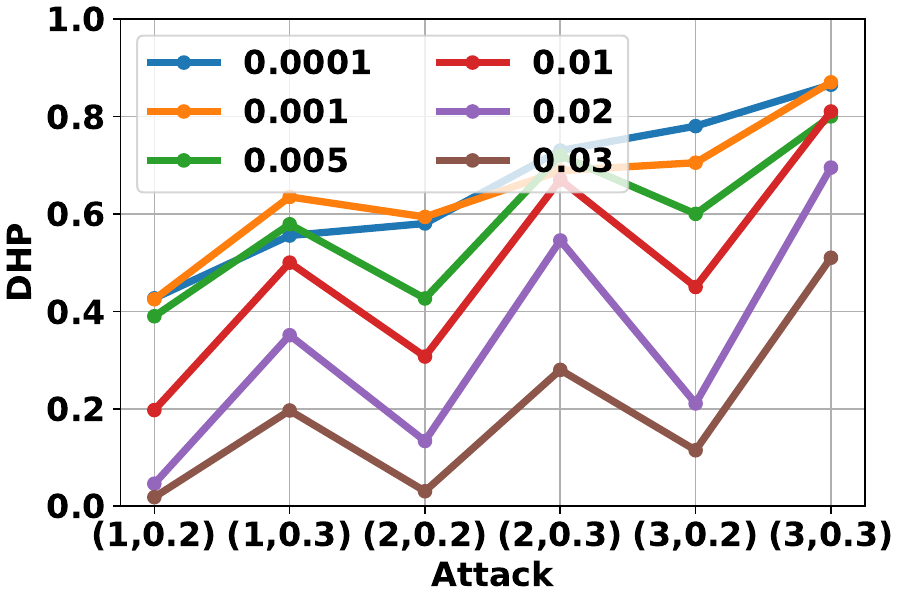}
         \caption{}
     \end{subfigure}
        \caption{Evaluation on different $\varrho$s: (a). MTD effectiveness; (b). MTD hiddenness.}
        \label{fig:va}
\end{figure}

First, based on the identification accuracy in Fig. \ref{fig:iden}, the effectiveness and hiddenness of MTD are summarised in Fig. \ref{fig:va}(a) and (b), respectively. In Fig. \ref{fig:va}(a), a larger attack is more likely to be detected and, in general, ADP increases and then decreases slightly as $\varrho$ increases. When $\varrho$ is small, the MTD is optimised on the limited set of candidate attack vectors around the identified attack, which may not include the actual attack vector. On the contrary, when $\varrho$ is large, the robust MTD is conservative by maximising the detection rate on the worst possible attack in a larger set, causing the actual detection rate to decrease. An extreme example is that when $\varrho > \|\bm{c}\|_2$, a zero-state attack vector becomes the worst-case attack, leading to a trivial solution to \eqref{eq:mtd_stage_one}. Regarding the hiddenness of MTD, Fig. \ref{fig:va}(b) shows that MTDs on a strong attack result in high DHP, which implies the trade-off between hiddenness and effectiveness. Referring to \eqref{eq:mtd_bilevel}, when the attack is strong, the effectiveness constraint can be more easily achieved, which in turn gives a lower residual of the attacker. Meanwhile, DHP decreases as $\varrho$ increases, which can be explained by a similar reason. 
In the following, $\varrho = 0.01$ will be used as the uncertainty bound in $\mathcal{C}$ due to its high ADP and moderate DHP.

\subsubsection{Comparison with Max-Rank and Robust MTDs}

\begin{figure}
     \centering
     \begin{subfigure}[b]{0.49\linewidth}
         \centering
         \includegraphics[width=1.00\linewidth]{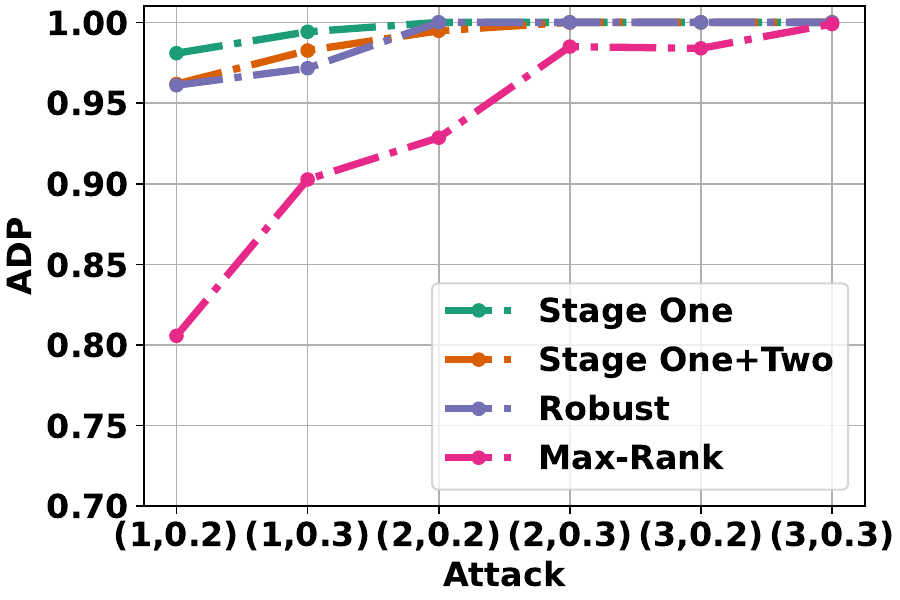}
         \caption{}
     \end{subfigure}
     \hfill
     \begin{subfigure}[b]{0.49\linewidth}
         \centering
         \includegraphics[width=\linewidth]{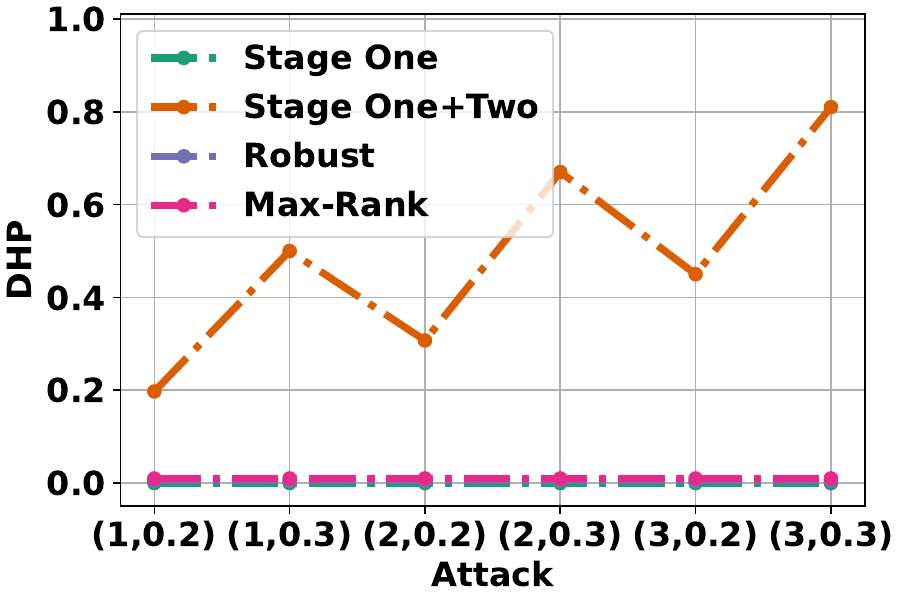}
         \caption{}
     \end{subfigure}
        \caption{Comparison on Stage-One, Stage-One + Stage-Two in DDET-MTD, Max-Rank MTD, and Robust MTD, assuming that the attack is detected by the LSTM-AE detector: (a). MTD effectiveness; (b) MTD hiddenness.}
        \label{fig:comparison}
\end{figure}

In this section, Stage-One and Stage-One + Stage-Two of the proposed DDET-MTD are compared with the Max-Rank MTD and Robust MTD algorithms. To fairly verify the performance of the proposed algorithm, both Max-Rank and Robust MTDs are triggered by the same LSTM-AE detector. Therefore, only the attacks that can be detected by the LSTM-AE are evaluated by the MTD algorithms, and we will leave a full comparison in the next section.
First, as shown in Fig. \ref{fig:comparison}(a), both Stage-One and Stage-One + Stage-Two can achieve ADPs greater than 96\% in all attack cases. The ADP of Stage-One is slightly higher than that of Stage-One + Stage-Two when the attack strength is small. This is because Stage-One maximises the residual, while Stage-Two adds the threshold as constraint. The proposed algorithm has an ADP comparable to the Robust MTD, which is significantly higher than the Max-Rank MTD. Therefore, it can be concluded that the event-triggered MTD does not significantly compromise the performance of the LSTM-AE detector shown in Fig. \ref{fig:roc}. 
Furthermore, Fig. \ref{fig:comparison}(b) shows that adding Stage-Two can significantly improve MTD hiddenness without compromising ADP. On the contrary, without considering the hiddenness of MTD, Stage-One, Robust and Max-Rank MTDs can always be detected by the attacker once the detector raises an alarm.


\subsection{Performances under Real-Time Operations}

In this section, more realistic power system operation is considered. The proposed DDET-MTD is compared with the Max-Rank MTD and the Robust MTD in both \textbf{periodic} and \textbf{event-triggering} settings. Meanwhile, as cyberattacks are very rare in practise, it is reasonable to discuss MTD usage and generator cost without attacks to see how the extra defence can impact normal system operations. In addition, the false positive rate reduction of LSTM-AE detector is also discussed.

\color{black}

\begin{table}
\centering
\caption{Average MTD performance under different attacks (in \%). The $\uparrow$ and $\downarrow$ represent the desired values being large and small, respectively. The best and second best performances are highlighted in red and blue, respectively. }
\footnotesize
  \begin{tabular}{clrrrrr}
    \hline\hline
    \multirow{2}{*}{\textbf{Attack}} & \multirow{2}{*}{\textbf{Metric}} &
      \multicolumn{2}{c}{\textbf{Periodic}} &
      \multicolumn{3}{c}{\textbf{Event-Triggered}} \\
    &  & {\textbf{Max}} & {\textbf{Robust}} & {\textbf{Max}} & {\textbf{Robust}} & {\textbf{DDET}} \\\hline\hline
    
    \multirow{4}{*}{(1,0.2)} 
    & $\uparrow$ ADP & 71.90 & \textcolor{red}{90.00} & 65.10 & 74.00 & \textcolor{blue}{75.50}  \\
    & $\uparrow$ DHP & 0.00  & 0.00  & 19.20 & \textcolor{blue}{23.00} & \textcolor{red}{41.24}  \\
    & $\downarrow$ Cost & 0.17  & 0.64  & \textcolor{blue}{0.14}  & 0.47  & \textcolor{red}{0.02} \\
    & $\downarrow$ Reac. & 27.53 & 45.69 & \textcolor{blue}{22.30} & 34.84 & \textcolor{red}{17.15} \\\hline
    
    \multirow{4}{*}{$(1,0.3)$} 
    & $\uparrow$ ADP & 83.40 & \textcolor{red}{93.00} & 79.70 & 84.50 & \textcolor{blue}{85.50}  \\
    & $\uparrow$ DHP & 0.00 & 0.00 & 11.70 & \textcolor{blue}{12.00} & \textcolor{red}{64.00}  \\
    & $\downarrow$ Cost & 0.17 & 0.61 & \textcolor{blue}{0.16} & 0.54 & \textcolor{red}{0.01} \\
    & $\downarrow$ Reac. & 27.73 & 45.50 & \textcolor{blue}{24.12} & 40.21 & \textcolor{red}{11.01} \\\hline
    
    \multirow{4}{*}{$(2,0.2)$} 
    & $\uparrow$ ADP & 93.10 & \textcolor{red}{98.50} & 90.81 & \textcolor{blue}{98.00} & 95.50  \\
    & $\uparrow$ DHP & 0.00 & 0.00 & \textcolor{blue}{2.20} & 2.00 & \textcolor{red}{34.72}  \\
    & $\downarrow$ Cost & 0.17 & 0.60 & \textcolor{blue}{0.16} & 0.60 & \textcolor{red}{0.01} \\
    & $\downarrow$ Reac. & 27.34 & 45.82 & \textcolor{blue}{26.95} & 44.63 & \textcolor{red}{16.56} \\\hline
    
    \multirow{4}{*}{$(2,0.3)$} 
    & $\uparrow$ ADP & \textcolor{blue}{96.91} & \textcolor{red}{100.00} & 98.10 & \textcolor{red}{100.00} & \textcolor{red}{100.00}  \\
    & $\uparrow$ DHP & 0.00 & 0.00 & \textcolor{blue}{0.40} & 0.00 & \textcolor{red}{67.00}  \\
    & $\downarrow$ Cost & \textcolor{blue}{0.17} & 0.64 & \textcolor{blue}{0.17} & 0.631 & \textcolor{red}{0.00} \\
    & $\downarrow$ Reac. & 27.45 & 45.63 & \textcolor{blue}{23.38} & 45.54 & \textcolor{red}{9.12} \\\hline
    
    \multirow{4}{*}{$(3,0.2)$} 
    & $\uparrow$ ADP & 98.90 & \textcolor{blue}{99.00} & 97.61 & \textcolor{blue}{99.00} & \textcolor{red}{100.00}  \\
    & $\uparrow$ DHP & 0.00 & 0.00 & 0.80 & \textcolor{blue}{1.00} & \textcolor{red}{45.00} \\
    & $\downarrow$ Cost & 0.18 & 0.62 & \textcolor{blue}{0.17} & 0.64 & \textcolor{red}{0.00} \\
    & $\downarrow$ Reac. & \textcolor{blue}{27.24} & 45.21 & 27.36 & 45.28 & \textcolor{red}{13.66} \\\hline
    
    \multirow{4}{*}{$(3,0.3)$} 
    & $\uparrow$ ADP & 99.80 & \textcolor{red}{100.00} & \textcolor{blue}{99.90} & \textcolor{red}{100.00} & \textcolor{red}{100.00}  \\
    & $\uparrow$ DHP & 0.00 & 0.00 & 0.00 & 0.00 & \textcolor{red}{81.00}  \\
    & $\downarrow$ Cost & \textcolor{blue}{0.17} & 0.66 & 0.18 & 0.65 & \textcolor{red}{0.00}\\
    & $\downarrow$ Reac. & \textcolor{blue}{27.53} & 45.54 & 27.57 & 45.85 & \textcolor{red}{7.33} \\\hline\hline
    
  \end{tabular}
  \label{tab:ave_compare}
\end{table}

\subsubsection{Operations under FDI Attack}
Average performances of Max-Rank MTD (Max), Robust MTD (Robust), and DDET-MTD (DDET) under different attack scenarios are compared in Table \ref{tab:ave_compare}. 

In general, DDET-MTD has the highest DHPs under each attack. Note that the DHPs of event-triggered Max-Rank and Robust MTDs are not zero due to the missing alarms (false negative samples) from the LSTM-AE detector. The false negative rate of LSTM-AE detector also causes the lower ADP of DDET-MTD than the periodic Robust MTD when the attack strength is low (see Fig. \ref{fig:roc}(b)). However, the periodic Robust MTD is the least economical method and cannot improve the hiddenness of MTD.

Thanks to the attack uncertainty set $\mathcal{C}$, the DDET-MTD can detect the attack with fewer efforts, resulting in the best economic performance of the lowest average reactance perturbation. Additionally, when the attack strength increases, the reactance perturbation ratio decreases, which can save the usage of D-FACTS devices in real-time operation. In contrast, as Robust MTD considers the worst detection performance all the time, it has the worst economic performance. The Robust and Max-Rank MTDs have almost constant average ratios per D-FACTS device under both periodic and event-triggering settings, as both algorithms cannot reflect different attack strengths and can easily over-react most of the time.

Although optimisation \eqref{eq:mtd_abstract} does not take the generator cost into account, simulation shows that DDET-MTD results in the lowest cost increase under each attack for two reasons. First, the DDET-MTD has the minimum reactance deviation against the default reactance settings. Therefore, its operational point is the closest to the optimal setting. Second, by improving the MTD hiddenness, the pre- and post- MTD power flows become similar to each other, resulting in less flow redistribution and similar line losses. Furthermore, Table \ref{tab:ave_compare} also illustrates that smaller costs are needed to detect more intense attacks in DDET-MTD, which is similar to the reactance perturbation.

\begin{figure}[h]
     \centering
     \begin{subfigure}[b]{0.49\linewidth}
         \centering
         \includegraphics[width=0.98\linewidth]{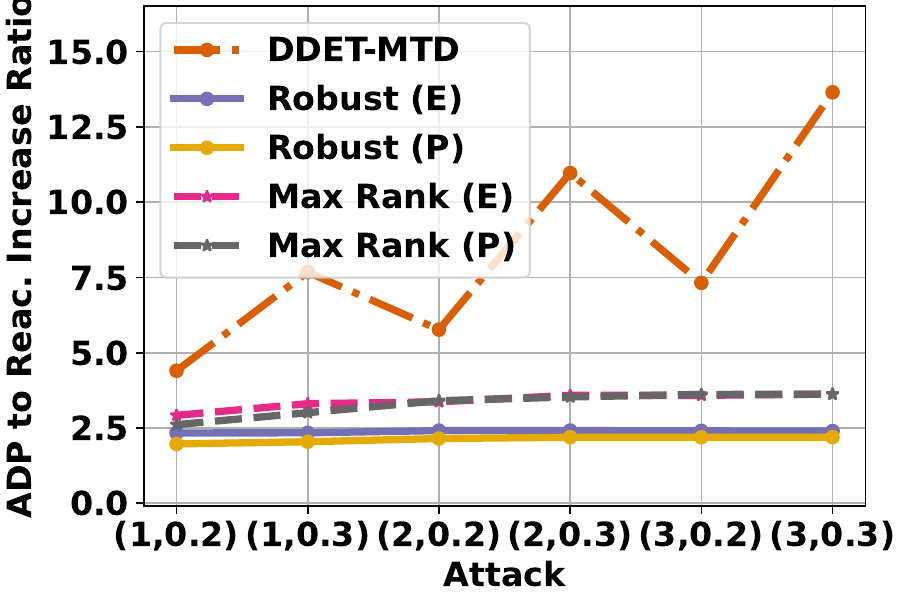}
         \caption{}
     \end{subfigure}
     \hfill
     \begin{subfigure}[b]{0.49\linewidth}
         \centering
         \includegraphics[width=0.98\linewidth]{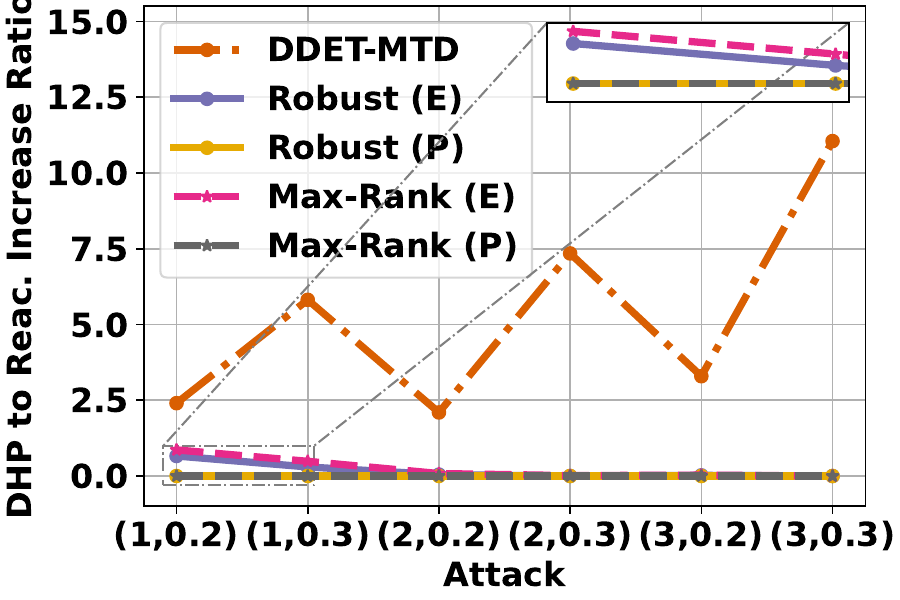}
         \caption{}
     \end{subfigure}
        \caption{a). Average ADP to average reactance increase ratio; (b). Average DHP to average reactance perturbation ratio. (E) and (P) represent the event-triggered and periodic settings, respectively.}
        \label{fig:ADP_DHP_X}
\end{figure}

To better illustrate the performances, Fig. \ref{fig:ADP_DHP_X} calculates the ratio of ADP and DHP with respect to the average perturbation ratio. It can be demonstrated that the DDET-MTD has the best trade-off between attack defence and operation economics, especially when the attack strength is high.

\color{black}

\subsubsection{False Positive Rejection}

\begin{figure}[h]
     \centering
     \begin{subfigure}[b]{0.49\linewidth}
         \centering
         \includegraphics[width=1.0\linewidth]{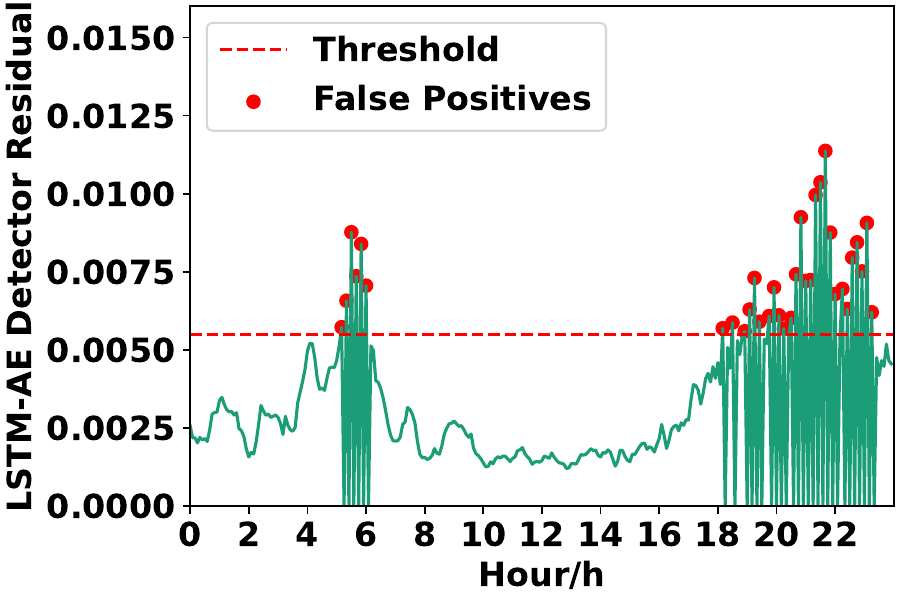}
         \caption{}
     \end{subfigure}
     \hfill
     \begin{subfigure}[b]{0.49\linewidth}
         \centering
         \includegraphics[width=1.0\linewidth]{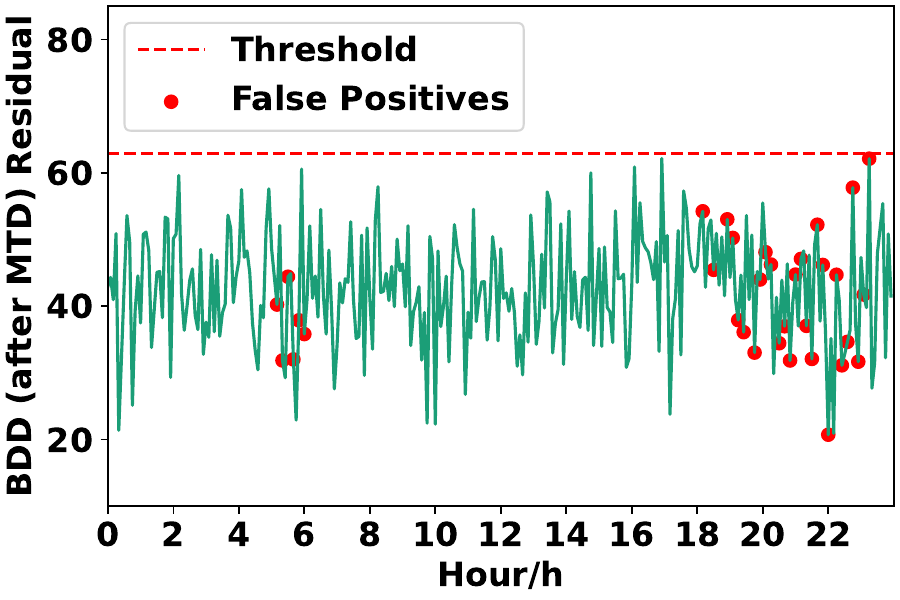}
         \caption{}
     \end{subfigure}
        \caption{False positive rejection on LSTM-AE detector using event-triggered MTD: (a). Residual of LSTM-AE detector; and (b). Residual of BDD (possibly after MTD).}
        \label{fig:rejection}
\end{figure}

Fig. \ref{fig:rejection}(a) records the residuals of the LSTM-AE detector in a single day from the test set. Positive samples are highlighted as red circles. There are many false positive alarms during the night, which can be caused by irregular use of electricity. Once the LSTM-AE detector raises an alarm, the attack identification and MTD will be triggered. As there is no ongoing attack, the residual of the post-MTD measurement follows the $\chi^2$ distribution. Consequently, as shown by Fig. \ref{fig:rejection}(b), all the false positive samples have residuals lower than the BDD threshold and no further actions are needed by the system operator. On average, the FPR of the LSTM-AE detector is reduced from 12.84\% to 1.84\% on test set after applying DDET-MTD. Note that the MTD FPR is well controlled by the predetermined BDD FPR $\alpha=2.0\%$.

\subsubsection{Normal Operations}

We now compare the economic performances of different MTD strategies without FDI attacks. The results in Table \ref{tab:normal} demonstrate that event triggering can significantly reduce the reactance perturbation and extra operational cost of MTDs. Meanwhile, the proposed DDET-MTD has the least interference with normal system operation, which makes it a promising defence against rare FDI attacks.


\color{black}


\begin{table}[h]
\centering
\caption{Average economical performances under normal operations (in \%)}
\footnotesize
  \begin{tabular}{lrrrrr}
    \hline\hline
     \multirow{2}{*}{\textbf{Metric}} &
      \multicolumn{2}{c}{\textbf{Periodic}} &
      \multicolumn{3}{c}{\textbf{Event-Triggered}} \\
    & {\textbf{Max}} & {\textbf{Robust}} & {\textbf{Max}} & {\textbf{Robust}} & {\textbf{DDET}} \\\hline\hline

    $\downarrow$ Cost & 0.171  & 0.628  & \textcolor{blue}{0.019}  & 0.059  & \textcolor{red}{0.005} \\
    $\downarrow$ Reac. & 27.470 & 45.565 & \textcolor{blue}{3.159} & 5.069 & \textcolor{red}{2.334}
     \\\hline\hline
  \end{tabular}
  \label{tab:normal}
\end{table}

\section{Conclusion}\label{sec:conclusion}

This paper proposes a novel Data-Driven Event-Triggered MTD algorithm to achieve high TPR and low FPR against FDI attacks, which can benefit both the data-driven detector and the physics-based MTD. Numeric simulations verify that the proposed DDET-MTD has better defence and economic trade-off compared with two baseline algorithms. On one hand, the high FPR of data-driven detector (12.8\%) is reduced by the MTD with controllable FPR (1.8\%) during the normal operation. During the attack, the attack identification serves as a bridge between data and physics. The roughly identified attack vector informs to design a bilevel hidden and effective MTD algorithm, which is further solved via two-stage convex relaxations. Thanks to the knowledge of the grid model, the DDET-MTD has a comparable detection accuracy as the robust MTD (96\%) while improves the hiddenness by 50\%. On the other hand, the inevitable extra cost of operating MTD on the physical power grid is negligible through the triggering mechanism and optimal design enabled by the data-driven detector. The proposed DDET-MTD can significantly reduce the reactance perturbation by 70\%, compared to the Robust MTD.

\appendix
\numberwithin{equation}{section}
\setcounter{equation}{0}

\subsection{Convert the Range of Reactance into Susceptance}
\label{sec:app_susceptance}

Let the branch $i$ have resistance $\bm{r}_i\in(0,\infty)$ and reactance $\bm{x}_i\in(0,\infty)$. The susceptance $\bm{b}_i$ can be determined as:
\begin{equation*}
    \bm{b}_i(\bm{x}_i) = \frac{-\bm{x}_i}{\bm{r}_i^2 + \bm{x}_i^2}
\end{equation*}

Therefore, $\bm{b}_i$ decreases monotonically on $(0,\bm{r}_i)$ and increases on $(\bm{r}_i,\infty)$. Considering the permissible range of $\bm{x}_i\in[\bm{x}_i^{-},\bm{x}_i^{+}]$, the permissible range of $\bm{b}_i$ can be determined. For $\bm{r}_i\in[\bm{x}_i^{-},\bm{x}_i^{+}]$, $\bm{b}^{-} = \frac{-\bm{r}_i}{\bm{r}_i^2+\bm{x}_i^2}$, $\bm{b}^{+} = \max{\left(\frac{-\bm{x}_i^+}{\bm{r}_i^2+{\bm{x}_i^{+}}^2}, \frac{-\bm{x}_i^-}{\bm{r}_i^2+{\bm{x}_i^{-}}^2} \right)}$; for $\bm{r}_i\notin[\bm{x}_i^{-},\bm{x}_i^{+}]$, $\bm{b}^{-} = \min{\left(\frac{-\bm{x}_i^+}{\bm{r}_i^2+{\bm{x}_i^{+}}^2}, \frac{-\bm{x}_i^-}{\bm{r}_i^2+{\bm{x}_i^{-}}^2} \right)}$, $\bm{b}^{+} = \max{\left(\frac{-\bm{x}_i^+}{\bm{r}_i^2+{\bm{x}_i^{+}}^2}, \frac{-\bm{x}_i^-}{\bm{r}_i^2+{\bm{x}_i^{-}}^2} \right)}$.



\color{black}

\subsection{Proof to Proposition \ref{theorem:mtd_stage_one}}
\label{sec:app_1}

To start, the Schur complement \cite{gallier2010notes} is given as follows.

\begin{theorem}\label{theorem: schur_1}
Given any symmetric matrix $\bm{Z}=\begin{bmatrix}\bm{A} & \bm{B}\\\star & \bm{C}\end{bmatrix}$, if $\bm{C}$ is invertible, the following two conditions are equivalent: (1) If $\bm{C}\succ 0$, then $\bm{Z}\succeq0$; (2) $\bm{A}-\bm{B}\bm{C}^{-1}\bm{B}^T\succeq0$.
\end{theorem}

\begin{proposition}\label{theorem: schur_2}
Given any symmetric matrix $\bm{Z}=\begin{bmatrix}\bm{A} & \bm{B}\\\star & \bm{C}\end{bmatrix}$, the following two conditions are equivalent: (1) $\bm{Z}\succeq0$; (2) $\bm{C}\succeq0$, $(\bm{I}-\bm{C}\bm{C}^\dagger)\bm{B}^T=0$, $\bm{A}-\bm{B}\bm{C}^\dagger\bm{B}^T\succeq0$.
\end{proposition}

First, the inner problem of \eqref{eq:mtd_stage_one} can be written as:
\begin{equation}\label{eq:inner}
    \begin{aligned}
    \min & \quad \|\bm{S}_0'\bm{H}_1\bm{c}'\|_2^2 \\
    \text{subject to} & \quad \|\bm{c}'-\bar{\bm{c}}\|_2^2\leq \varrho^2
    \end{aligned}
\end{equation}
where $\bm{S}_0' = \mathcal{S}(\bm{H}_0')$ and the Lagrangian of \eqref{eq:inner} is written as:
\begin{equation}\label{eq:lagrangian}
    \mathcal{L}(\bm{c}',\nu) = {\bm{c}'}^T\left(\bm{H}_1^T\bm{S}_0'\bm{H}_1+\nu\bm{I}\right)\bm{c'}+(-2\nu\bar{\bm{c}}^T)\bm{c}'+\nu(\bar{\bm{c}}^T\bar{\bm{c}}-\varrho^2)
\end{equation}

Based on \eqref{eq:lagrangian} and denoting $\bm{M}=\bm{H}_1^T\bm{S}_0'\bm{H}_1+\nu\bm{I}$, the dual function of \eqref{eq:inner} can be analytically written as
\begin{equation}
\begin{array}{l}\label{eq:mtd_stage_one_dual_function}
    g(\nu) = \inf_{\bm{c}'}\mathcal{L}(\bm{c}',\nu) \\
    \;  =
     \begin{cases}
            \begin{array}{l}
                -(\nu\bar{\bm{c}})^T\bm{M}^\dagger(\nu\bar{\bm{c}})
                +\nu(\bar{\bm{c}}^T\bm{\bar{c}}-\varrho^2)
            \end{array} 
             & \bm{M}\succeq0, \text{ and} \\
            & \quad \nu\bar{\bm{c}}\in\text{Col}(\bm{M})  \\
      \ -\infty & \text{otherwise}
    \end{cases}
\end{array}
\end{equation}

Let $-(\nu\bar{\bm{c}})^T\bm{M}^\dagger(\nu\bar{\bm{c}})+\nu(\bar{\bm{c}}^T\bm{\bar{c}}-\varrho^2)\geq\omega$. The dual problem of \eqref{eq:inner} becomes:
\begin{equation}\label{eq:inner_dual}
    \begin{aligned}
    \max_{\nu,\omega} & \quad \omega \\
    \text{subject to} & \quad \nu \geq 0 \\
    & \quad \nu(\bar{\bm{c}}^T\bm{\bar{c}}-\varrho^2) - \omega - (\nu\bar{\bm{c}})^T\bm{M}^\dagger(\nu\bar{\bm{c}}) \geq 0 \\
    & \quad \bm{M} \succeq 0 \\
    & \quad \nu\bar{\bm{c}} \in \text{Col}(\bm{M})
    \end{aligned}
\end{equation}

Note that the last constraint of \eqref{eq:inner_dual} can be rewritten as $\bm{M}\bm{M}^\perp\nu\bar{\bm{c}} = \nu\bar{\bm{c}}$. Applying Proposition \ref{theorem: schur_2}, the dual problem can be rewritten as
\begin{equation}\label{eq:inner_dual_1}
    \begin{aligned}
    \max_{\nu,\omega} & \quad \omega \\
    \text{subject to} & \quad \nu \geq 0 \\
    & \quad \begin{bmatrix}
        \nu(\bar{\bm{c}}^T\bar{\bm{c}}-\varrho^2) - \omega & (\nu\bar{\bm{c}})^T \\
        \star & \bm{M}
    \end{bmatrix} \succeq 0
    \end{aligned}
\end{equation}

The strong duality between \eqref{eq:inner} and \eqref{eq:inner_dual_1} holds as long as $\mathcal{C}\neq \emptyset $ \cite{boyd2004convex}. Consequently, the robust stage one problem \eqref{eq:mtd_stage_one} becomes:
\begin{subequations}\label{eq:mtd_stage_1_dual}
    \begin{alignat}{2}
    \max_{\bm{b}',\mu,\omega} & \quad \omega \\
    \text{subject to} 
    & \quad [\bm{b}'] - [\bm{b}^-] \succeq 0,  [\bm{b}^+] - [\bm{b}'] \succeq 0 \label{eq:mtd_stage_1_dual_dfacts} \\
    & \quad \nu \geq 0 \label{eq:mtd_stage_1_dual_variable}\\
    & \quad \begin{bmatrix}
        \nu(\bar{\bm{c}}^T\bar{\bm{c}}-\varrho^2) - \omega & (\nu\bar{\bm{c}})^T \label{eq:mtd_stage_1_dual_lmi_1}\\
        \star & \bm{M}
    \end{bmatrix} \succeq 0
    \end{alignat}
\end{subequations}

Note that $\bm{M}=\nu\bm{I}+\bm{H}_1^T\bm{H}_1- \bm{H}_1^T\bm{H}_0'(\bm{H}_0^{'T}\bm{H}_0')^{-1}\bm{H}_0^{'T}\bm{H}_1$ is nonlinear in the decision variable $\bm{b}'$. In Theorem \ref{theorem: schur_1}, define $\bm{A}:=\begin{bmatrix}
        \nu(\bar{\bm{c}}^T\bar{\bm{c}}-\varrho^2) - \omega & (\nu\bar{\bm{c}})^T \\
        \star & \nu\bm{I}+\bm{H}_1^T\bm{H}_1
    \end{bmatrix}$, $\bm{B}:=\begin{bmatrix} \bm{0} \\ \bm{H}_1^T\bm{H}_0' \end{bmatrix}$, and $\bm{C}: = \bm{H}_0^{'T}\bm{H}_0'$ in \eqref{eq:mtd_stage_1_dual_lmi_1}. Since $\bm{C}>0$ and Theorem \ref{theorem: schur_1}, the constraint \eqref{eq:mtd_stage_1_dual_lmi_1} becomes \eqref{eq:mtd_stage_one_lmi_matrix}, which finalises the proof.


\subsection{Proof to Proposition \ref{theorem:mtd_stage_one_ite}}
\label{sec:app_ite}
First, the following sufficient condition holds for any matrices $\bm{E},\bm{E}_0$ with the same dimension \cite{liu2015computation}:
\begin{equation*}
    \bm{E}_0^T\bm{E}+\bm{E}^T\bm{E}_0 - \bm{E}_0^T\bm{E}_0 \succeq 0 \Rightarrow \bm{E}^T\bm{E} \succeq 0
\end{equation*}

Define $\bm{E} = \bm{C}^N+\bm{V}^N[\bm{b}']\bm{A}_r^c$ and $\bm{E}_0 = \bm{V}^N[\bm{b}_k]\bm{A}_r^c$. Replacing ${\bm{H}_0'}^{T}\bm{H}_0'$ in \eqref{eq:mtd_stage_one_lmi_matrix} by $\bm{H}_{\text{update}} = \bm{E}_0^T\bm{E}+\bm{E}^T\bm{E}_0 - \bm{E}_0^T\bm{E}_0$ finalises the proof.

\subsection{Proof to Proposition \ref{theorem:mtd_stage_two}}
\label{sec:app_2}

The dual function \eqref{eq:mtd_stage_one_dual_function} is the lower bound of the primary function, e.g. $g(\nu)\leq \|\bm{S}_0'\bm{H}_1\bm{c}'\|_2^2$ for $\forall \bm{c}'\in\mathcal{C}$. Therefore, a sufficient condition for \eqref{eq:mtd_bilevel_nest} is $g(\nu)\geq\omega$. Note that $\lambda_c(\rho)$ is replaced by constant $\omega$ in stage two. Therefore, Proposition \ref{theorem:mtd_stage_two} can be proved similarly to Proposition \ref{theorem:mtd_stage_one}. Furthermore, define the cost of \eqref{eq:mtd_bilevel} as $(\bm{b}'-\bm{b}_0)^T\bm{H}_{\text{hid}}^T\bm{I}\bm{H}_{\text{hid}}(\bm{b}'-\bm{b}_0) \leq \phi$. Applying Theorem \ref{theorem: schur_1} on $\bm{I}$ gives \eqref{eq:mtd_stage_two_new}.

\subsection{Hidden and Effective MTD Algorithm}
\label{sec:app_algorithm}

The hidden and effective MTD algorithm is summarised in Algorithm \ref{alg:mtd} in detail. The inputs of the algorithm $\mathcal{B}$, $\mathcal{C}$, $\bm{C}^N$, $\bm{V}^N$, $\bm{A}_r^c$, $\bm{H}_1$, $\bm{H}_{\text{hid}}$, and $\lambda_c'$ have been defined in the main content. Further, define $tol_{\text{one}}$ and $tol_{\text{two}}$ as the tolerance of stage-one and stage-two problem, respectively. Define $ite_{\text{one}}$ and $ite_{\text{two}}$ as the maximum iteration step in stage-one and stage-two problem, respectively. Meanwhile, let $no$ be the multi-run number. The output of this algorithm is the set-point of the D-FATCS devices, denoted as $\bm{b}_{\text{mtd}}$.


\begin{algorithm}[t]
    \footnotesize
    \SetKwInOut{Input}{Input}
    \SetKwInOut{Output}{Output}

    \Input{$\mathcal{B}$, $\mathcal{C}$, $\bm{C}^N$, $\bm{V}^N$, $\bm{A}_r^c$, $\bm{H}_1$, $\bm{H}_{\text{hid}}$, $\lambda_c'$, $tol_{one}$, $tol_{two}$, $ite_{one}$, $ite_{two}$, $no$}
    \Output{$\bm{b}_{\text{mtd}}$}
    
    
    \tcc{Stage-One Algorithm}
    
    $\mathcal{D}^{\text{one}}=\{\cdot\}$, $\mathcal{G}^{\text{one}} = \{\cdot\}$ \tcc{Store multi-run results}
    
    $i = 0$
    
    \While{$i \leq no$}
    {
    
    $k=0$, $\omega_{k} = 0$
    
    Random generate $\bm{b}_k \in \mathcal{B}$
    
    \While{$k \leq ite_{one}$}
    {
    Solve \eqref{eq:mtd_stage_one_iteration}. Record the optimal value as $\omega^\star$ and optimal solution as $\bm{b}'$
    
    $\bm{b}_{k+1} \leftarrow \bm{b}'$
    
    \If{$\omega^\star - \omega_{k} \leq tol_{one}$}
    {
    break
    }
    
    $k\leftarrow k+1$, $\omega_{k} = \omega^\star$
    
    }
    
    $\mathcal{D}^{\text{one}} = \{\mathcal{D}^{\text{one}},\bm{b}'\}$, $\mathcal{G}^{\text{one}} = \{\mathcal{G}^{\text{one}},\omega^\star\}$
    
    }
    
    \tcc{Stage-Two Algorithm}
    
    $\omega^* = \max{\mathcal{G}}$
    
    \eIf{$\lambda_c' > \omega^*$}
    {
    Define $\mathcal{I}^{\text{two}} = \{i|\omega_i = \arg\max{\mathcal{G}^{\text{one}}}\}$
    
    $\mathcal{D}^{\text{two}} = \{\mathcal{D}^{\text{one}}[i] | i\in \mathcal{I}^{\text{two}}\}$
    
    $\omega = \omega^*$
    
    }
    {
    Define $\mathcal{I}^{\text{two}} = \{i|\omega_i \geq \lambda_c', \omega_i\in\mathcal{G}^{\text{one}}\}$
    
    $\mathcal{D}^{\text{two}} = [\mathcal{D}^{\text{one}}[i] | i\in\mathcal{I}^{\text{two}}]$
    
    $\omega = \lambda_c'$
    }
    
    $\mathcal{P}=\{\cdot\}$, $\mathcal{Q} = \{\cdot\}$
    
    \For{$\bm{b}\in\mathcal{D}^{\text{two}}$}
    {
    
    $k=0$, $\bm{b}_k = \bm{b}$, $\phi_k=1e+5$
    
    \While{$k \leq ite_{two}$}
    {
    
    Solve \eqref{eq:mtd_stage_two_iteration}. Record the optimal value as $\phi^\star$ and optimal solution as $\bm{b}'$
    
    \If{$\phi_{k}  - \phi^\star\leq tol_{two}$}
    {
    break
    }
    
    $k\leftarrow k+1$, $\phi_{k} = \phi^\star$
    
    }
    
    $\mathcal{P} = \{\mathcal{P},\bm{b}'\}$, $\mathcal{Q} = \{\mathcal{Q},\phi^\star\}$
    
    }
    
    $\bm{b}_{\text{mtd}} = \mathcal{P}[\arg\min_i{\mathcal{Q}}]$
    
    \caption{Hidden and Effective MTD Algorithm}
    \label{alg:mtd}
\end{algorithm}

\color{black}

\bibliographystyle{IEEEtran}
\bibliography{IEEEabrv,Reference.bib}

\begin{thebibliography}{10}
\providecommand{\url}[1]{#1}
\csname url@samestyle\endcsname
\providecommand{\newblock}{\relax}
\providecommand{\bibinfo}[2]{#2}
\providecommand{\BIBentrySTDinterwordspacing}{\spaceskip=0pt\relax}
\providecommand{\BIBentryALTinterwordstretchfactor}{4}
\providecommand{\BIBentryALTinterwordspacing}{\spaceskip=\fontdimen2\font plus
\BIBentryALTinterwordstretchfactor\fontdimen3\font minus
  \fontdimen4\font\relax}
\providecommand{\BIBforeignlanguage}[2]{{%
\expandafter\ifx\csname l@#1\endcsname\relax
\typeout{** WARNING: IEEEtran.bst: No hyphenation pattern has been}%
\typeout{** loaded for the language `#1'. Using the pattern for}%
\typeout{** the default language instead.}%
\else
\language=\csname l@#1\endcsname
\fi
#2}}
\providecommand{\BIBdecl}{\relax}
\BIBdecl

\bibitem{mahmoud2021cyberphysical}
M.~S. Mahmoud, H.~M. Khalid, and M.~M. Hamdan, \emph{Cyberphysical
  Infrastructures in Power Systems: Architectures and Vulnerabilities}.\hskip
  1em plus 0.5em minus 0.4em\relax Academic Press, 2021.

\bibitem{khalid2016bayesian}
H.~M. Khalid and J.~C.-H. Peng, ``A bayesian algorithm to enhance the
  resilience of wams applications against cyber attacks,'' \emph{IEEE
  Transactions on Smart Grid}, vol.~7, no.~4, pp. 2026--2037, 2016.

\bibitem{hug2012vulnerability}
G.~Hug and J.~A. Giampapa, ``Vulnerability assessment of ac state estimation
  with respect to false data injection cyber-attacks,'' \emph{IEEE Transactions
  on Smart Grid}, vol.~3, no.~3, pp. 1362--1370, 2012.

\bibitem{deng2017false}
R.~Deng, G.~Xiao, R.~Lu, H.~Liang, and A.~V. Vasilakos, ``False data injection
  on state estimation in power systems—attacks, impacts, and defense: A
  survey,'' \emph{IEEE Transactions on Industrial Informatics}, vol.~13, no.~2,
  pp. 411--423, 2017.

\bibitem{bellizio2022transition}
F.~Bellizio, W.~Xu, D.~Qiu, Y.~Ye, D.~Papadaskalopoulos, J.~L. Cremer, F.~Teng,
  and G.~Strbac, ``Transition to digitalized paradigms for security control and
  decentralized electricity market,'' \emph{Proceedings of the IEEE}, pp.
  1--18, 2022.

\bibitem{cheng2022highly}
G.~Cheng, Y.~Lin, J.~Zhao, and J.~Yan, ``A highly discriminative detector
  against false data injection attacks in ac state estimation,'' \emph{IEEE
  Transactions on Smart Grid}, 2022.

\bibitem{higgins2021topology}
M.~Higgins, J.~Zhang, N.~Zhang, and F.~Teng, ``Topology learning aided false
  data injection attack without prior topology information,'' in \emph{2021
  IEEE Power Energy Society General Meeting (PESGM)}, 2021, pp. 1--5.

\bibitem{rahman2014moving}
M.~A. Rahman, E.~Al-Shaer, and R.~B. Bobba, ``Moving target defense for
  hardening the security of the power system state estimation,'' in
  \emph{Proceedings of the First ACM Workshop on Moving Target Defense}, 2014,
  pp. 59--68.

\bibitem{liu2018reactance}
C.~Liu, J.~Wu, C.~Long, and D.~Kundur, ``Reactance perturbation for detecting
  and identifying fdi attacks in power system state estimation,'' \emph{IEEE
  Journal of Selected Topics in Signal Processing}, vol.~12, no.~4, pp.
  763--776, 2018.

\bibitem{zhang2019analysis}
Z.~Zhang, R.~Deng, D.~K. Yau, P.~Cheng, and J.~Chen, ``Analysis of moving
  target defense against false data injection attacks on power grid,''
  \emph{IEEE Transactions on Information Forensics and Security}, vol.~15, pp.
  2320--2335, 2019.

\bibitem{liu2020optimal}
B.~Liu and H.~Wu, ``Optimal d-facts placement in moving target defense against
  false data injection attacks,'' \emph{IEEE Transactions on Smart Grid},
  vol.~11, no.~5, pp. 4345--4357, 2020.

\bibitem{higgins2022cyber}
M.~Higgins, W.~Xu, F.~Teng, and T.~Parisini, ``Cyber-physical risk assessment
  for false data injection attacks considering moving target defences,''
  \emph{International Journal of Information Security}, 2022.

\bibitem{lakshminarayana2020cost}
S.~Lakshminarayana and D.~K. Yau, ``Cost-benefit analysis of moving-target
  defense in power grids,'' \emph{IEEE Transactions on Power Systems}, vol.~36,
  no.~2, pp. 1152--1163, 2020.

\bibitem{liu2022explicit}
M.~Liu, C.~Zhao, Z.~Zhang, and R.~Deng, ``Explicit analysis on effectiveness
  and hiddenness of moving target defense in ac power systems,'' \emph{IEEE
  Transactions on Power Systems}, pp. 1--1, 2022.

\bibitem{zhang2022voltage}
H.~Zhang, B.~Liu, X.~Liu, A.~Pahwa, and H.~Wu, ``Voltage stability constrained
  moving target defense against net load redistribution attacks,'' \emph{IEEE
  Transactions on Smart Grid}, vol.~13, no.~5, pp. 3748--3759, 2022.

\bibitem{xu2022robust}
W.~Xu, I.~M. Jaimoukh, and F.~Teng, ``Robust moving target defence against
  false data injection attacks in power grids,'' \emph{IEEE Transactions on
  Information Forensics and Security}, pp. 1--1, 2022.

\bibitem{tian2018enhanced}
J.~Tian, R.~Tan, X.~Guan, and T.~Liu, ``Enhanced hidden moving target defense
  in smart grids,'' \emph{IEEE transactions on smart grid}, vol.~10, no.~2, pp.
  2208--2223, 2018.

\bibitem{zhang2020hiddenness}
Z.~Zhang, R.~Deng, D.~K. Yau, P.~Cheng, and J.~Chen, ``On hiddenness of moving
  target defense against false data injection attacks on power grid,''
  \emph{ACM Transactions on Cyber-Physical Systems}, vol.~4, no.~3, pp. 1--29,
  2020.

\bibitem{liu2021optimal}
B.~Liu and H.~Wu, ``Optimal planning and operation of hidden moving target
  defense for maximal detection effectiveness,'' \emph{IEEE Transactions on
  Smart Grid}, vol.~12, no.~5, pp. 4447--4459, 2021.

\bibitem{higgins2021enhanced}
M.~Higgins, K.~Mayes, and F.~Teng, ``Enhanced cyber-physical security using
  attack-resistant cyber nodes and event-triggered moving target defence,''
  \emph{IET Cyber-Physical Systems: Theory \& Applications}, vol.~6, no.~1, pp.
  12--26, 2021.

\bibitem{xu2022physical}
W.~Xu, I.~M. Jaimoukha, and F.~Teng, ``Physical verification of data-driven
  cyberattack detector in power system: An mtd approach,'' in \emph{2022 IEEE
  PES Innovative Smart Grid Technologies Conference Europe (ISGT-Europe)},
  2022, pp. 1--5.

\bibitem{wu2021extreme}
T.~{Wu}, W.~{Xue}, H.~{Wang}, C.~Y. {Chung}, G.~{Wang}, J.~{Peng}, and
  Q.~{Yang}, ``Extreme learning machine-based state reconstruction for
  automatic attack filtering in cyber physical power system,'' \emph{IEEE
  Transactions on Industrial Informatics}, vol.~17, no.~3, pp. 1892--1904,
  2021.

\bibitem{zhang2021detecting}
Y.~{Zhang}, J.~{Wang}, and B.~{Chen}, ``Detecting false data injection attacks
  in smart grids: A semi-supervised deep learning approach,'' \emph{IEEE
  Transactions on Smart Grid}, vol.~12, no.~1, pp. 623--634, 2021.

\bibitem{xu2020deep}
W.~Xu and F.~Teng, ``A deep learning based detection method for combined
  integrity-availability cyber attacks in power system,'' \emph{arXiv preprint
  arXiv:2011.01816}, 2020.

\bibitem{pang2021deep}
G.~Pang, C.~Shen, L.~Cao, and A.~V.~D. Hengel, ``Deep learning for anomaly
  detection: A review,'' \emph{ACM Computing Surveys (CSUR)}, vol.~54, no.~2,
  pp. 1--38, 2021.

\bibitem{ahmed2020challenges}
C.~M. Ahmed, G.~R. MR, and A.~P. Mathur, ``Challenges in machine learning based
  approaches for real-time anomaly detection in industrial control systems,''
  in \emph{Proceedings of the 6th ACM on Cyber-Physical System Security
  Workshop}, 2020, pp. 23--29.

\bibitem{ashok2018online}
A.~{Ashok}, M.~{Govindarasu}, and V.~{Ajjarapu}, ``Online detection of stealthy
  false data injection attacks in power system state estimation,'' \emph{IEEE
  Transactions on Smart Grid}, vol.~9, no.~3, pp. 1636--1646, 2018.

\bibitem{zimmerman2018matpower}
R.~Zimmerman and C.~MurilloSanchez, ``Matpower user’s manual version 7.0
  b1,'' \emph{Power Systems Engineering Research Center (PSerc): Tempe, AZ,
  USA}, 2018.

\bibitem{abur2004power}
A.~Abur and A.~G. Exposito, \emph{Power system state estimation: theory and
  implementation}.\hskip 1em plus 0.5em minus 0.4em\relax CRC press, 2004.

\bibitem{zhang2021smart}
H.~Zhang, B.~Liu, and H.~Wu, ``Smart grid cyber-physical attack and defense: A
  review,'' \emph{IEEE Access}, vol.~9, pp. 29\,641--29\,659, 2021.

\bibitem{gao2022novel}
S.~Gao, J.~Lei, X.~Wei, Y.~Liu, and T.~Wang, ``A novel bilevel false data
  injection attack model based on pre-and post-dispatch,'' \emph{IEEE
  Transactions on Smart Grid}, pp. 1--1, 2022.

\bibitem{kingma2014adam}
D.~P. Kingma and J.~Ba, ``Adam: A method for stochastic optimization,''
  \emph{arXiv preprint arXiv:1412.6980}, 2014.

\bibitem{liu2021interior}
B.~Liu, Q.~Yang, H.~Zhang, and H.~Wu, ``An interior-point solver for ac optimal
  power flow considering variable impedance-based facts devices,'' \emph{IEEE
  Access}, vol.~9, pp. 154\,460--154\,470, 2021.

\bibitem{liu2019joint}
C.~Liu, H.~Liang, T.~Chen, J.~Wu, and C.~Long, ``Joint admittance perturbation
  and meter protection for mitigating stealthy fdi attacks against power system
  state estimation,'' \emph{IEEE Transactions on Power Systems}, vol.~35,
  no.~2, pp. 1468--1478, 2019.

\bibitem{sinha2017review}
A.~Sinha, P.~Malo, and K.~Deb, ``A review on bilevel optimization: from
  classical to evolutionary approaches and applications,'' \emph{IEEE
  Transactions on Evolutionary Computation}, vol.~22, no.~2, pp. 276--295,
  2017.

\bibitem{liu2021optimalcoding}
C.~Liu, R.~Deng, W.~He, H.~Liang, and W.~Du, ``Optimal coding schemes for
  detecting false data injection attacks in power system state estimation,''
  \emph{IEEE Transactions on Smart Grid}, vol.~13, no.~1, pp. 738--749, 2021.

\bibitem{wang2021multi}
J.~Wang, W.~Xu, Y.~Gu, W.~Song, and T.~Green, ``Multi-agent reinforcement
  learning for active voltage control on power distribution networks,''
  \emph{Advances in Neural Information Processing Systems}, vol.~34, 2021.

\bibitem{paszke2019pytorch}
A.~Paszke, S.~Gross, F.~Massa, A.~Lerer, J.~Bradbury, G.~Chanan, T.~Killeen,
  Z.~Lin, N.~Gimelshein, L.~Antiga \emph{et~al.}, ``Pytorch: An imperative
  style, high-performance deep learning library,'' \emph{Advances in neural
  information processing systems}, vol.~32, 2019.

\bibitem{diamond2016cvxpy}
S.~Diamond and S.~Boyd, ``{CVXPY}: {A} {P}ython-embedded modeling language for
  convex optimization,'' \emph{Journal of Machine Learning Research}, vol.~17,
  no.~83, pp. 1--5, 2016.

\bibitem{gallier2010notes}
\BIBentryALTinterwordspacing
J.~H. Gallier. (2010) Notes on the schur complement. [Online]. Available:
  \url{https://repository.upenn.edu/cgi/viewcontent.cgi?article=1637&context=cis_papers}
\BIBentrySTDinterwordspacing

\bibitem{boyd2004convex}
S.~Boyd and L.~Vandenberghe, \emph{Convex optimization}.\hskip 1em plus 0.5em
  minus 0.4em\relax Cambridge university press, 2004.

\bibitem{liu2015computation}
C.~Liu and I.~M. Jaimoukha, ``The computation of full-complexity polytopic
  robust control invariant sets,'' in \emph{2015 54th IEEE Conference on
  Decision and Control (CDC)}, 2015, pp. 6233--6238.

\end{thebibliography}

\bstctlcite{IEEEexample:BSTcontrol}


\begin{IEEEbiography}[{\includegraphics[width=1in,height=1.25in,clip,keepaspectratio]{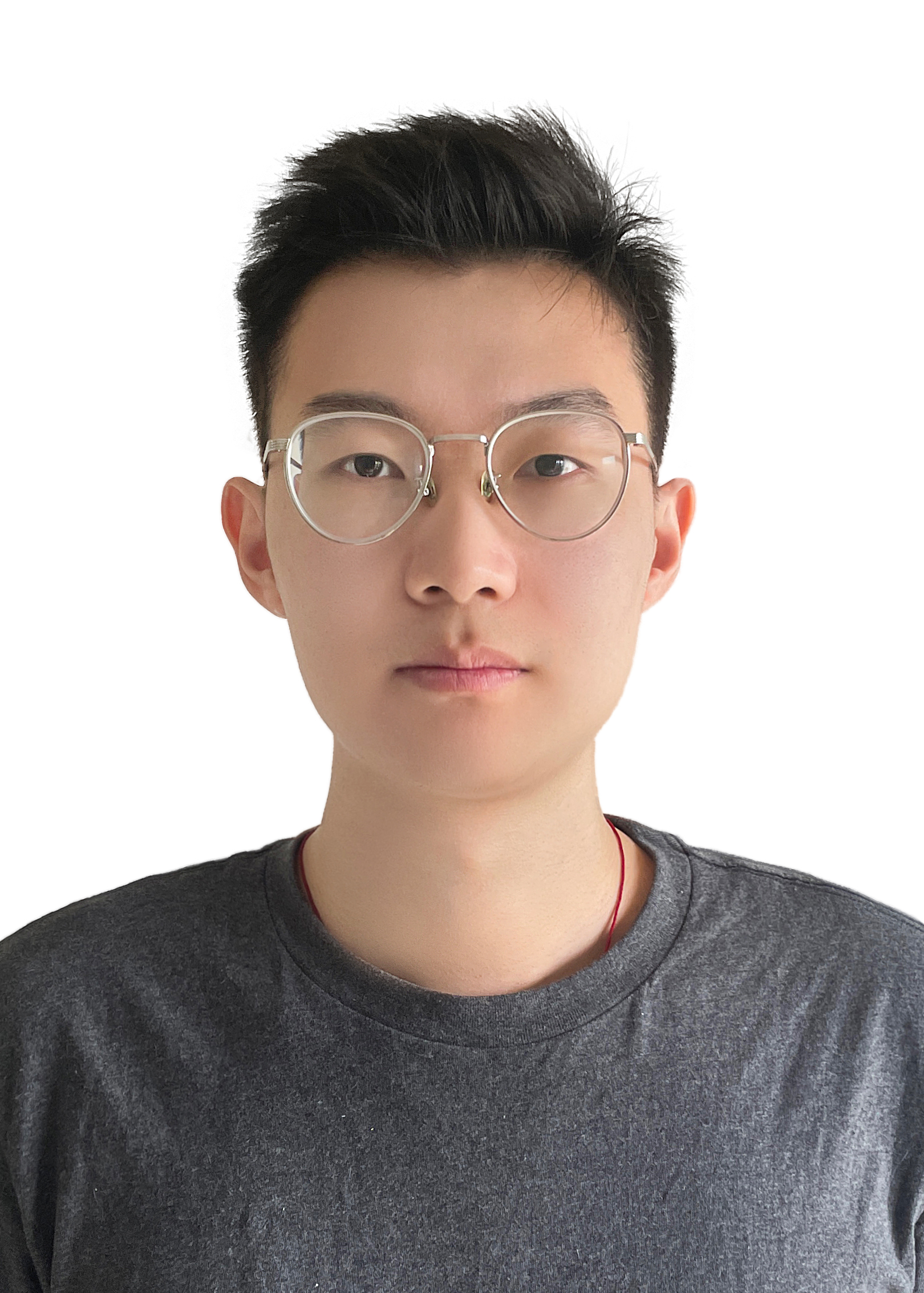}}]{Wangkun Xu} (Student Member, IEEE) received B.Eng. degree in electrical and electronic engineering from University of Liverpool, UK in 2018 and M.Sc. degree in control systems from Imperial College London, UK in 2019, where he is currently a Ph.D. student. His research focuses on robust and privacy-preserving machine learnings in cyber-physical power system operation and security.
\end{IEEEbiography}

\begin{IEEEbiography}[{\includegraphics[width=1in,height=1.25in,clip,keepaspectratio]{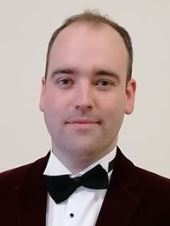}}]{Martin Higgins} (Member, IEEE) received his BSc in Physics from Queen Mary, University of London, in 2011 and his MSc and Ph.D. from Imperial College London, U.K. in 2012 and 2022 respectively. His Ph.D. thesis was on False Data Injection attacks against power systems and was achieved as part of the CDT in Smart Grids integrated MRES/PHD. Currently, Martin is a research associate at the University of Oxford contributing on the Digital Security by Design Project.  His research interests include power systems cyber-security, self-driving vehicles, sensor spoofing attacks, false data injection attacks and moving target defence.
\end{IEEEbiography}

\begin{IEEEbiography}[{\includegraphics[width=1in,height=1.25in,clip,keepaspectratio]{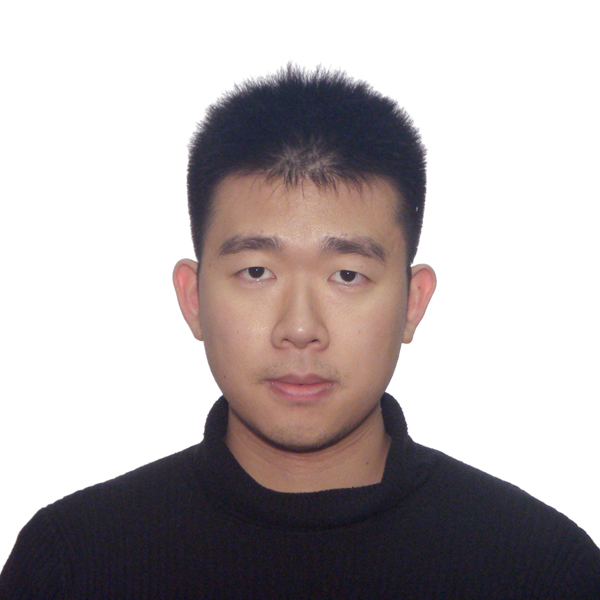}}]{Jianhong Wang} received B.Eng. degree in Computer Science and Electronic Engineering from University of Liverpool, UK in 2016, M.Sc. degree in Computing (Machine Learning) from Imperial College London, UK in 2017, and M.Res. degree in Web Science and Data Analytics from University College London, UK in 2018. He is currently pursuing the Ph.D. degree in Electrical Engineering Research at Imperial College London, UK. His research interests lie in multi-agent reinforcement learning and its applications to the real-world problems. He has published several papers in AI top conferences such as AAAI, NeurIPS and ICLR.
\end{IEEEbiography}

\begin{IEEEbiography}[{\includegraphics[width=1in,height=1.25in,clip,keepaspectratio]{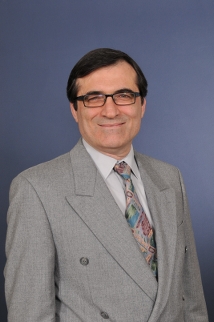}}]{Imad M. Jaimoukha} received the B.Sc. degree in electrical engineering from the University of Southampton, Southampton, U.K., in 1983, and the M.Sc. and Ph.D. degrees in control systems from Imperial College London, London, U.K., in 1986 and 1990, respectively. He was a Research Fellow with the Centre for Process Systems Engineering at ICL from 1990 to 1994. Since 1994, he has been with the Department of Electrical and Electronic Engineering, ICL. His research interests include robust and fault-tolerant control, system approximation, and global optimization.
\end{IEEEbiography}

\begin{IEEEbiography}[{\includegraphics[width=1in,height=1.25in,clip,keepaspectratio]{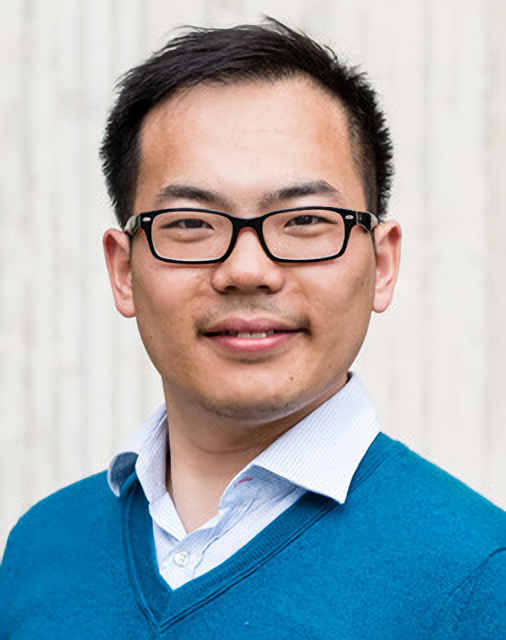}}]{Fei Teng} (Senior Member, IEEE) received the B.Eng. degree in electrical engineering from Beihang University, China, in 2009, and the M.Sc. and Ph.D. degrees in electrical engineering from Imperial College London, U.K., in 2010 and 2015, respectively, where he is currently a Senior Lecturer with the Department of Electrical and Electronic Engineering. His research focuses on the power system operation with high penetration of Inverter-Based Resources (IBRs) and the Cyber-resilient and Privacy-preserving cyber-physical power grid.
\end{IEEEbiography}

\ifCLASSOPTIONcaptionsoff
  \newpage
\fi
\end{document}